 \documentclass[preprint2,longabstract]{aastex}

\newcommand{\oiii}{[O~{\sc iii}] }
\newcommand{\nii}{[N~{\sc ii}] }

\slugcomment{Submitted to the Astronomical Journal}

\shorttitle{Fading AGN host galaxies }
\shortauthors{Keel et al.}

\begin{document}


\title{HST Imaging of Fading AGN Candidates I: Host-Galaxy Properties and Origin of the Extended Gas\footnote{Based on observations with the 
NASA/ESA {\it Hubble Space Telescope} obtained at the Space Telescope Science Institute, which is operated
by the Association of Universities for Research in Astronomy, Inc.,
under NASA contract No. NAS5-26555.}}


\author{William C. Keel \altaffilmark{1,2}}
\affil{Department of Physics and Astronomy, University of Alabama, Box 870324, Tuscaloosa, AL 35487}
\email{wkeel@ua.edu, @NGC3314}

\author{W. Peter Maksym}
\affil{Department of Physics and Astronomy, University of Alabama, Box 870324, Tuscaloosa, AL 35487}
\email{wpmaksym@ua.edu, @StellarBones}

\author{Vardha N. Bennert}
\affil{Physics Department, California Polytechnic State University,
San Luis Obispo, CA 93407}
\email{vbennert@calpoly.edu}

\author{Chris J. Lintott}
\affil{Astrophysics, Oxford University; and Adler Planetarium, 1300 S. Lakeshore Drive, Chicago, IL 60605}
\email{cjl@astro.ox.ac.uk; @chrislintott}

\author{S. Drew Chojnowski}
\affil{Department of Astronomy,	 New Mexico State University, P. O. Box 30001, MSC 4500, Las Cruces, New Mexico 88003-8001}

\author{Alexei Moiseev}
\affil{Special Astrophysical Observatory, Russian Academy of Sciences, Nizhny Arkhyz, Russia 369167}

\author{Aleksandrina Smirnova}
\affil{Special Astrophysical Observatory, Russian Academy of Sciences, Nizhny Arkhyz, Russia 369167}

\author{Kevin Schawinski}
\affil{Institute for Astronomy, ETH Z\"urich,
Wolfgang-Pauli-Strasse 27,
CH-8093 Zurich,
Switzerland}
\email{kevin.schawinski@phys.ethz.ch; @kevinschawinski}

\author{C. Megan Urry}
\affil{Department of Physics,
Yale University,
P.O. Box 208120,
New Haven, CT 06520-8120}

\author{Daniel A. Evans}
\affil{Harvard-Smithsonian Center for Astrophysics, 60 Garden Street,
MS-67, Cambridge, MA 02138}

\author{Anna Pancoast}
\affil{Department of Physics, University of California, Santa Barbara, CA 93106}


\author{Bryan Scott}
\affil{Physics Department, California Polytechnic State University,
San Luis Obispo, CA 93407}

\author{Charles Showley}
\affil{Physics Department, California Polytechnic State University,
San Luis Obispo, CA 93407}

\and 

\author{Kelsi Flatland}
\affil{Physics Department, California Polytechnic State University,
San Luis Obispo, CA 93407}


\altaffiltext{1}{Visiting astronomer, Kitt Peak National Observatory, National Optical Astronomy Observatories, which is operated by the 
Association of Universities for Research in Astronomy, Inc. (AURA) under cooperative agreement with the National Science 
Foundation.}
\altaffiltext{2}{SARA Observatory}


\begin{abstract}
We present narrow- and medium-band HST imaging, with additional supporting
ground-based imaging, spectrophotometry, and Fabry-Perot interferometric data, for eight galaxies identified as hosting fading AGN. These are selected to have AGN-ionized gas projected $>10$ kpc from the nucleus, and energy budgets with a significant shortfall
of ionizing radiation between the requirement to ionize the distant gas and the AGN as observed directly, indicating fading of the 
AGN on $\approx 50,000$-year timescales.
This paper focuses on the host-galaxy properties and origin of the gas.  In every galaxy, we identify evidence of ongoing or past interactions, including
tidal tails, shells, and warped or chaotic dust structures; a similarly-selected sample of obscured AGN with extended ionized clouds shares this 
high incidence of disturbed morphologies. Several systems show multiple dust lanes in different orientations, broadly fit by
differentially precessing disks of accreted material viewed $\sim 1.5$ Gyr after its initial arrival. The host systems are of early Hubble type; most show 
nearly pure de Vaucouleurs 
surface-brightness profiles and Sersic indices appropriate for classical bulges, with one S0 and one SB0 galaxy. The gas has systematically lower
metallicity than the nuclei; three systems have abundances uniformly well below solar, consistent with an origin in tidally disrupted low-luminosity
galaxies, while some systems have more nearly solar abundances (accompanied by such signatures as multiple Doppler components), 
which may suggest redistribution of gas by outfows within the host galaxies themselves. 
These aspects are consistent with a tidal origin for the extended gas in most systems, although the ionized gas and stellar tidal features
do not always match closely. Unlike extended emission regions around many radio-loud AGN, these clouds are kinematically dominated
by rotation, in some cases in warped disks. Outflows can play important kinematic roles only in localized regions near some of the AGN.
We find only a few sets of young star clusters potentially triggered
by AGN outflows. In UGC 7342 and UGC 11185, multiple luminous star clusters are seen just within the projected ionization cones, potentially marking star formation triggered by outflows. As in the discovery example Hanny's Voorwerp/IC 2497, there are regions in these clouds where lack of a strong correlation between H$\alpha$ 
surface brightness and ionization parameter indicates that there is unresolved fine structure in the clouds. Together with thin coherent filaments spanning several kpc, 
persistence of these structures over their orbital lifetimes may require a role for magnetic confinement. Overall, we find that the sample of fading AGN occurs
in interacting and merging systems, that the very extended ionized gas is composed of tidal debris rather than galactic winds, and that these host systems
are bulge-dominated and show no strong evidence of triggered star formation in luminous clusters.
\end{abstract}




\section{Introduction}

An important subset of active galactic nuclei (AGN) is the small fraction with extended emission-line regions (EELRs), ionized gas
extending as far as tens of kpc from the core. These regions provide information on the geometry of escaping ionizing radiation,
(as in ionization cones), on the history of the AGN via light-travel time to the clouds, and on the environment of the AGN. In radio-loud
objects, these regions often have complex and large-amplitude velocity structures, indicating an important role for outflows. In contrast,
radio-quiet objects with EELRs may show lower-amplitude and globally ordered velocity fields, consistent with tidal debris photoionized by the AGN but not
otherwise disturbed.

Recent results have highlighted an important role for EELRs in probing the history of AGN radiative output on timescales up to $10^5$ years.
An early discovery of the Galaxy Zoo citizen-science project \citep{cjl2008} was Hanny's Voorwerp, a galaxy-sized cloud of
highly-ionized gas associated with the spiral galaxy IC 2497 \citep{cjl2009}. The object's spectrum clearly indicates photoionization by
a luminous AGN, while direct observations of IC 2497 at all wavelengths show a deficit of at least two orders of magnitude in ionizing luminosity
(\citealt{cjl2009}, \citealt{kevin2010}). Furthermore, HST imaging and spectroscopy fail to show any high-ionization gas in IC 2497 which would
result from an extant AGN \citep{HSTHV}. These results are all explained if the AGN in IC 2497 was a QSO until $\approx 10^5$ years before
our current view of it, and we see Hanny's Voorwerp as an ionization echo - the analog of a light echo, but expressed through recombination 
which has long timescales at low densities. Such dramatic changes in radiative output suggest correspondingly dramatic changes in the accretion
onto the central black hole, providing a view on hitherto inaccessible timescales. Using the properties of Hanny's Voorwerp as a template, and
enabled by nearly 200 Galaxy Zoo volunteers, we searched for similar objects, using the distinct colors of nebulae with strong \oiii $\lambda 5007$ emission
when seen at low redshift in the SDSS filter set \citep{mnras2012}; by analogy with Hanny's Voorwerp, these were known collectively as ``Voorwerpjes" using the Dutch diminutive form. 

A spectroscopic study of the most promising candidates from color and morphology
yielded a sample of 19 additional AGN with high-ionization gas extending more than 10 kpc in projection from the AGN, suitable for studying the history of
these AGN. Among these, eight show a strong deficit in ionizing luminosity based on both the nuclear emission-line fluxes and far-infrared luminosity,
thereby excluding objects with strongly obscured AGN from which radiation escapes in some directions to ionize surrounding gas. 
Fading AGN are easier to find with this approach than brightening ones, since these have the same signature as a lack of surrounding cool
gas at an appropriate position to be ionized. Broadly, the relative numbers of fading versus obscured AGN in the \cite{mnras2012} sample
suggests that these AGN are luminous in episodes lasting $0.2 - 2 \times 10^5$ years, probably with many such episodes over much longer
lifetimes. At least 3/4 of these are in interacting or merging systems, again consistent with ionizing gas-rich tidal debris at large radii
in directions not blocked by a disk in the AGN host galaxy itself.
Subsequent work by other groups has extended the case for long-timescale variations in AGN. A set of objects with very similar properties
at higher redshifts and higher luminosities has been reported by  \cite{Schirmer},
and deep \oiii images of the merger remnant  NGC 7252 by \cite{Schweizer}
indicate a recent active episode of its nucleus (where there is no directly detected AGN at all). In a statistical context, \cite{Hickox2014} find that
similarly dramatic AGN variations can account for the scatter in relationships between star-formation indicators and AGN luminosity.

We have carried out an observational campaign to characterize the Voorwerpje systems in more detail, concentrating on the fading-AGN
candidates, aiming both to probe the history of the AGN itself and examine the properties of the host galaxies and their interaction and star-forming histories.
We concentrate in this paper on the structure of the host galaxies, origin and fine structure of the ionized gas, and potential evidence for
star formation triggered by outflow from the AGN.
Companion papers deal with the implications of these data for these AGN and their histories (Lintott et al., in preparation), and with the two sample AGN whose
double radio sources and giant ionized clouds violate the usual alignment pattern for radio jets and ionization cones (Mkn 1498 and NGC 5972;
Maksym et al., in preparation). The
high-resolution HST images support the interpretation that the AGN in these objects are indeed fading, give hints as to the timescales involved, and in
some cases suggest nuclear outflows possibly connected to the fading or mode change of the AGN. We describe in turn the various kinds of observations
we have carried out, the analysis techniques used in our measurements, and the results organized both by kind of data and (in an appendix) by galaxy.

In evaluating sizes and luminosities, we compute distances using a Hubble constant 
73 km s$^{-1}$ Mpc$^{-1}$.

\section{Sample, Observations, and Data Reduction}

\subsection{Galaxy sample}

We observed the subset of EELR host galaxies from \cite{mnras2012} which have ionizing-energy deficits -- ratios of bolometric AGN luminosity to
the required level for ionizing the most distant clouds -- greater than about a factor 3. (The companion paper by Lintott et al.  includes updated
energy-budget calculations, incorporating WISE data to fill mid-IR gaps in the SED). Table \ref{tbl-1} lists the target galaxies, AGN types, and redshifts.

\subsection{HST imaging}

Our primary data set consists of HST images isolating \oiii $\lambda 5007$ and H$\alpha$+\nii $\lambda 6583$ and
neighboring continuum bands. For the starlight continuum, we used the F621M (center 6190 {\AA}, FWHM =700 {\AA}) and F763M 
(center 7600 {\AA}, FWHM=780 {\AA}) intermediate-band filters of the Wide-Field Camera 3 (WFC3), 
to combine reasonable surface-brightness sensitivity with good rejection of the strong emission lines.
For the emission lines, we used the ramp filters on the Advanced Camera for Surveys (ACS) except for a few instances
where a galaxy's redshift put one of these lines in a narrow WFC3 filter. The ACS narrow ramp filters have nominal FWHM of 2\%
of the center wavelength.
Total exposures were 1250-1400 seconds in F621M, 1150-1260 seconds in F763M, and 2800-2900 seconds in the narrow
bands, with the variations depending on viewing period per orbit. The observation sequences were set to give slightly longer exposures
in H$\alpha$, known to be weaker than \oiii emission. Each observation included two exposures with small 0 \farcs 145 dither offsets
between them, to improve sampling of the final product and reduce effects of hot or low pixels.
For NGC 5972, already known to have very large emission regions, the dither 
step of 2 \farcs 41 was made large enough to fill the inter-chip gap with one exposure.
The different cameras often required different guide stars, leading
to small offsets ($\simeq 0 \farcs 2$) in the world coordinate systems between them; we used the nuclei and bright stars to register the sets of images.

The CCDs in both ACS and WFC3 are affected by charge-transfer 
efficiency (CTE) losses, especially severe in the low-background
regime of the narrowband data. Uncorrected, this leads to much of the
charge originating in a pixel being spread along its readout
column; not only does this lead to underestimating the flux of 
faint targets, but cosmic-ray events leave trails which are
harder to reject cleanly. These are substantially improved through
application of pixel-based reconstructions, essentially a 
deconvolution using physical properties of the CTE losses.
We used the ACS routine of \citet{AndersonBedin},
which is now incorporated in the MAST archival processing, and the recently-released WFC3 version
described by \cite{WFC3CTE}. The output of these correction routines
was combined for each filter using {\tt multidrizzle} \citep{drizzle}.
The {\tt LAcosmic} routine \citep{vDokkum} worked well 
to reject and patch cosmic-ray events for the WFC3
data; we used interactive removal with the IRAF {\tt imedit} task for ACS 
images.

Table \ref{tbl-1} lists the
filters used for the narrow emission-line bands for each galaxy. For brevity,
we refer to SDSS J151004.01+074037.1 as SDSS 1510+07,
to SDSS J220141.64+115124.3 as SDSS 2201+11, and note that
the Teacup AGN is formally known as SDSS J143029.88+133912.0.

For most objects, line/continuum separation is very clean with these
filters. For Mkn 1498 and the Teacup AGN, we used image arithmetic
to correct for line emission in the nominal continuum bands,
at levels of 10\% in narrowband count rate for \oiii and 30\% for
H$\alpha$+\nii. For Mkn 1498, the scale factors were 0.10 for \oiii 
and 0.15 for H$\alpha$+\nii. To improve continuum subtraction in \oiii
around the dust
lanes of SDSS 2201+11, we used a synthetic continuum image extrapolated from the observed
images for a closer match to the wavelength of the line emission, iteratively generated to minimize residuals in regions with weak or no line emission.

In the WFC3 images of UGC 7342, reflections from bright field stars produce bright elliptical rings 
superimposed on the outer parts of the host galaxy, in smooth regions so that little information is lost. 

NGC 5252, with its well-known ionization cones, satisfies our
sample criteria relating nuclear luminosity and ionization level in 
the outer regions. We use the archival HST WFPC2 images described by 
\cite{tsvetanov}  and \cite{morse},
with analysis parallel to our handling of the new ACS and WFC3 images.
The monochromatic field of the WFPC2 ramp filters is much smaller than the
40x80{\arcsec} field of the ACS ramps, so the outermost filaments are progressively
lost in the \oiii image.

For the two largest systems, NGC 5972 and UGC 7342, the telescope orientation for the ACS observations was constrained to align their major axes
nearly parallel to the orientation of the $40 \times 80${\arcsec} monochromatic field of view. The only known emission-line structure lost to the field size is 
an outlying, radial filament extending from 22-45{\arcsec} to the south of NGC 5972; ground-based data give independent information on its \oiii flux.

Fig. \ref{fig-overview1} gives an overview of the target galaxies in our HST imagery, comparing continuum structures and \oiii emission.

\begin{figure*} 
\includegraphics[width=148.mm,angle=0]{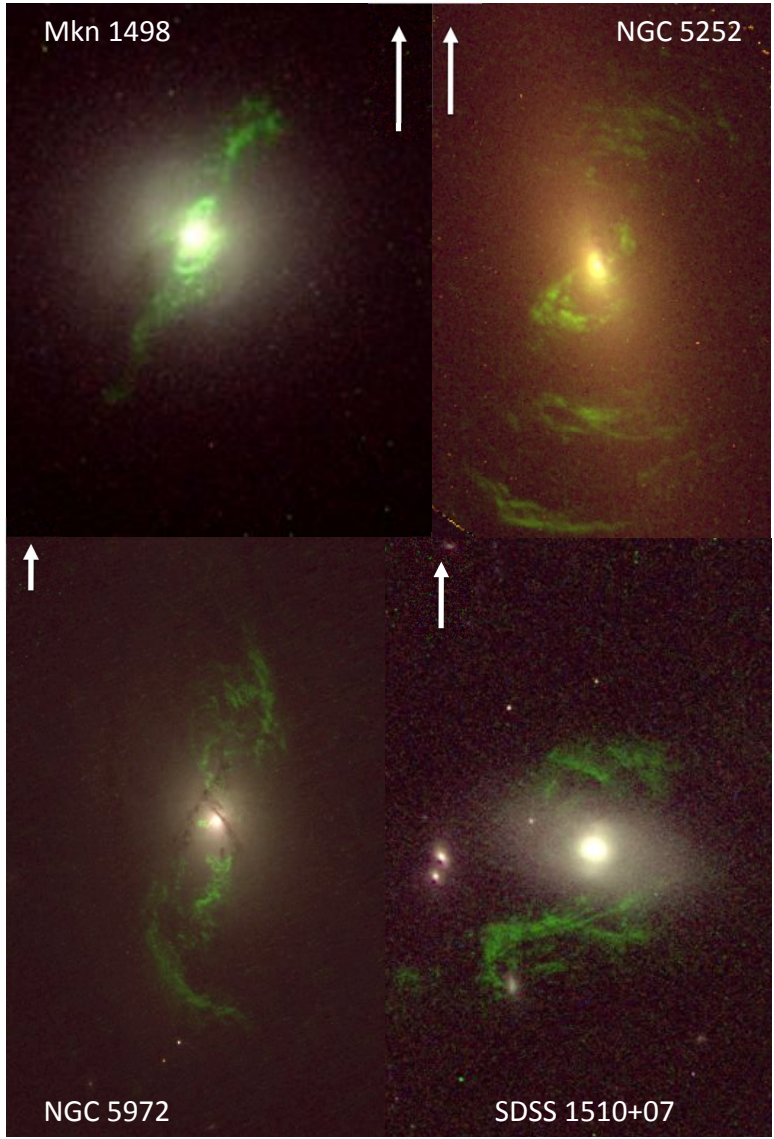} 
\caption{ Comparison of continuum and emission-line structure of the sample galaxies. Blue and red are the continuum WFC3 images, and
their average with \oiii emission (from ACS or WFC3 narrowband data) added is mapped to green. The scaling of \oiii to continuum is arbitrary but consistent for six objects;
for SDSS 1510+07 and SDSS 2201+11, with lower \oiii  surface brightness, we used a scaling three times
larger for the emission component. For each color band, we use an offset log intensity mapping (intensity $\propto \log ([I + {\rm constant}]$,
similar to the sinh mapping used in SDSS composite images). Intensity scales are consistent between images except for Mkn 1498,
where the maximum mapped intensity is twice as high as the rest to show detail near the AGN. Changes in the continuum color are
shown, but very small in this visualization due to the modest wavelength baseline between F621M and F763M filter bands. The angular scales are independently
chosen to show the emission structure; scale bars are 5{\arcsec} long with arrows pointing north. North is to the top and east to the left for all except UGC 7342.
 (Continued on next page)} 
\label{fig-overview1} 
\end{figure*}

\addtocounter{figure}{-1}

\begin{figure*} 
\includegraphics[width=163.mm,angle=0]{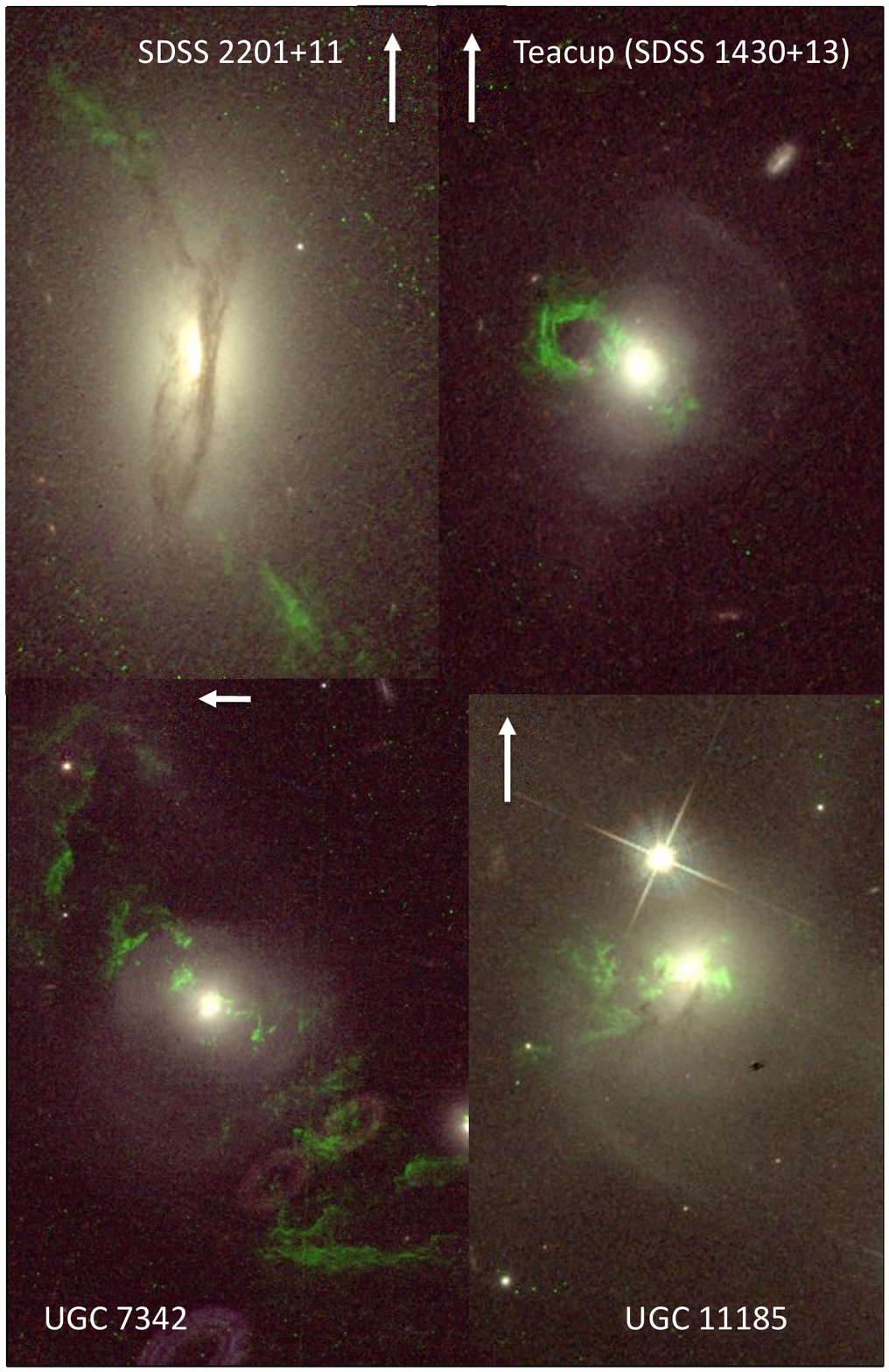} 
\caption{(continued)} 
\end{figure*} 

\begin{deluxetable}{lccccc}
\tablecaption{Galaxy properties and filters \label{tbl-1}}
\tablewidth{0pt}
\tablehead{
\colhead{Galaxy} &	\colhead{AGN	type} & \colhead{$z$} & \colhead{\oiii filter}  & \colhead{H$\alpha$+\nii filter} & \colhead{Scale, kpc/\arcsec}}
\startdata
Mkn 1498 &  Sy 1.9 & 0.0547 & 	ACS FR551N &	ACS FR716N	& 1.11 \cr
NGC 5252 & Sy 1.5 & 0.0228 &	WFPC2 FR533N &	WFPC2 F673N & 0.50 \cr
NGC 5972 & Sy 2 & 0.0297 & ACS FR505N &	WFC3 F673N & 0.63	\cr
SDSS 1510+07 & Sy 2 & 0.0458 &	ACS FR505N &	ACS FR716N	&  0.89 \cr
SDSS 2201+11 & Sy 2 & 0.0296 &	WFC3 FQ508N & 	WFC3 F673N	& 0.59  \cr
Teacup AGN & Sy 2 & 0.0852 &	ACS FR551N & 	ACS FR716N	&1.58  \cr
UGC 7342 & Sy 2 & 0.0477 & 	ACS FR505N &	ACS FR716N	& 0.98 \cr
UGC 11185 & Sy 2 & 0.0412 &	ACS FR505N &	ACS FR656N	& 0.80  \cr
\enddata
\end{deluxetable}

\subsubsection{Surface photometry \label{s-surfacephotometry}}

To assess their surface-brightness profiles and trace dust features, 
we fit families of elliptical isophotes to the continuum images using the 
{\it stsdas} {\tt ellipse} task (which implements the algorithm from \citealt{jedrzejewski})
and created smooth models following these profiles using 
the {\tt bmodel} task. The isophote-tracing task was run in an interactive 
mode allowing masking of obvious dust lanes or patches. In some regions 
where one of the ellipse parameters changes strongly with radius, the 
{\tt bmodel} output had arcs of missing values, which we replaced with the 
median  of nonzero values over $7 \times 7$-pixel regions.
For SDSS 2201+11, the dust is so widespread that our starting fit used the maximum intensity of the image after reflection about both axes relative to the large-scale major axis. Dust features were quantified from
attenuation images, the ratio between data and these models for each continuum image. Color-ratio images between the two continuum bands provide
an additional measure of the location and extent of dust attenuation,
although with lower signal-to-noise ratio because of the small wavelength
leverage between F621M and F763M passbands. The color-ratio image, however, is not subject to structure shortcomings of the elliptical-model fits in complex host galaxies, giving an independent view of the reality of
dust features.

\subsubsection{Identification and photometry of candidate star clusters \label{s-starclusters}}

Although the continuum bands were selected for their proximity to wavelengths of strong emission lines, we also
use them to characterize candidate star clusters. The color baseline between F621M and F763M is quite similar to
$R-I$ in the Bessell (essentially Cousins) system. We have estimated the transformations via synthetic photometry on the \cite{pickles} spectral library
using the response curves for these WFC3 filters and the $RI$ curves given by Pickles. For $R-I < 0.8$, there are closely
linear relations between the $m_{621} - m_{763}$ color and $R-I$; redder than this, for M stars, the F621M filter
includes TiO and CaH bands (e.g., \citealt{bochanski}), and the relation becomes nonlinear. We ignore this regime, since
our interest is in young and luminous star clusters. Using Vega magnitudes where colors for an A0 V stars are zero,
and incorporating  the median redshift $z=0.041$ of our objects,
we find $ m_{621}-m_{763} = 0.92 (R-I)_0 $ where the subscript denotes that $R-I$ is at redshift zero. The scatter about this
linear fit is 0.026 magnitude for all stars earlier than type M. The redshift correction is quite small in $ m_{621}-m_{763} $, 
introducing a scatter $\sigma = 0.013$ magnitude, and significant systematic trends only for M stars. Therefore, to estimate
standard colors of star clusters, we invert the above expression and use $(R-I)_0 = 1.09 ( m_{621}-m_{763} )$.
 
 In measuring magnitudes and colors of candidate star clusters, we normalize to a radius of $r=0.4${\arcsec} (8 pixels), using the
 modest curve-of-growth corrections for a point source (0.045 magnitude for F631M, 0.033 for F763M). We correct for
 foreground Milky Way extinction using the values from \cite{Schlafly}. To reduce the effects of gradients in the local background,
 we performed aperture photometry after subtracting a smooth galaxy model, and quote errors based on statistics of local background
 regions after this subtraction.
 
\subsection{Long-slit spectra}

We have obtained long-slit spectra of regions seen as particularly 
interesting in the HST images, using two facilities. At Lick Observatory, we 
used the Kast double spectrograph on the 3m Shane telescope. Separate red and blue detectors, on optical trains fed by a dichroic
beamsplitter, covered wavelength ranges 3400-4650 and 4900-7300 {\AA} . 
The slit width was 2\farcs 0.

At the 6-meter telescope (BTA) of the Special Astrophysical Observatory of 
the Russian Academy of Sciences, the SCORPIO-2 multi--mode focal reducer  \citep{AfanasievMoiseev} 
was used in the long--slit spectrograph mode, covering the wavelength 
range 3800--8100 {\AA} at a resolution typically 4.5  {\AA} FWHM. The slit width was varied with the
seeing, typically 1\farcs 0.  These spectra are detailed in Table \ref{tbl-2}.

For both sets of spectra, absolute flux calibration used observations of 
standard stars, within the scaling errors set by changes in atmospheric 
seeing. These data, along with the long-slit spectra described by
\cite{mnras2012}, provide information on the kinematics and chemical abundances of the gas
and constrain ionization mechanisms in various regions (as described in section \ref{sec-gaskin}, and Lintott et al., in preparation).

\begin{figure*} 
\includegraphics[width=145.mm,angle=90]{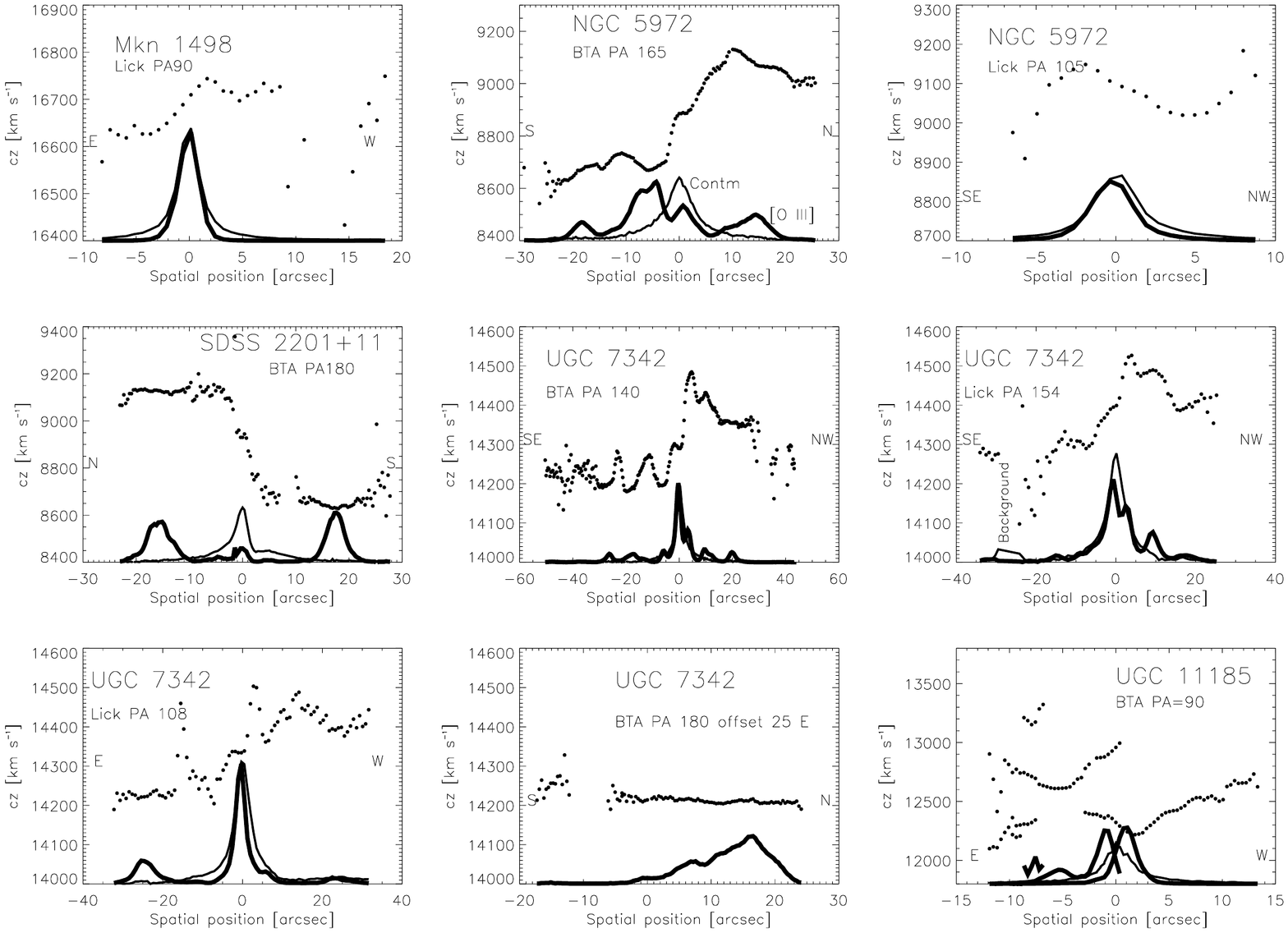} 
\caption{Velocity slices from the long-slit spectra in Table 2, derived from Gaussian fits to [O III] $\lambda 5007$ and shown as $cz$ in the heliocentric frame. For spatial comparison, traces of continuum and
[O III] flux along the slit are shown at the bottom of each panel, as in
Fig. 12 of \citep{mnras2012}.Their intensities are arbitrarily scaled for
visibility, with [O III] as the thicker lines. The offset spectrum of UGC 7342 does not have an associated continuum profile; 
at its distance from the core the continuum is
essentially undetected. The spectrum of UGC 7342 along PA $154^\circ$
crosses the background galaxy at $z=0.0674$, which appears as a continuum
enhancement at the indicated location.
UGC 11185 shows 3 distinct velocity systems; the gap
in the most extensive (low-velocity) system may be due to blending with the
locally stronger intermediate component. The [O III] profiles are shown 
separately for these components (at different intensity scales for 
visibility).} 
\label{fig-longslit} 
\end{figure*}

Some of the Lick spectra were targeted at possible outflow structures
near the nuclei, and lie near the minor axes of the extended emission.
The Lick PA $154^\circ$ spectrum includes the projected neighboring
galaxy  SDSS J121820.18+291447.7, which proved to be a background object at $z=0.0674$. For UGC 11185, the
BTA spectrum resolved three velocity systems in the emission lines
on the eastern side of the galaxy. The velocity results from these
spectra are summarized in Fig. \ref{fig-longslit}, combining the redshift values with
intensity slices for orientation with galaxy and emission-cloud
features.

\begin{deluxetable}{lccccc}
\tablecaption{Long-slit spectra \label{tbl-2}}
\tablewidth{0pt}
\tablehead{
\colhead{Galaxy} &	\colhead{PA\arcdeg} & \colhead{Telescope} & \colhead{Slit [\arcsec]} & \colhead{Exposure [s]} & \colhead{UT date} }
\startdata
NGC 5972 & 165 & BTA & 1.0 & 900 & 2013 April 6 \\
SDSS 2201+11 & 180 & BTA & 1.0 & 1800 & 2012  August 25\\
UGC 7342 & 180 & BTA & 0.7 & 2700 &  2013 April 5\\
UGC 7342 & 140 & BTA & 0.7 & 3600 & 2013 April 5 \\
UGC 11185 & 90 & BTA	& 1.0 & 1800& 2012 August 25\\
Mkn 1498	& 90 & Lick & 2.0 & 1800 & 2013 March 13\\
NGC 5972 & 105 & Lick & 2.0 & 1800  &  2013 January 18\\
UGC 7342 & 154 & Lick & 2.0 & 1800  & 2013 January 15 \\
UGC 7342 & 108 & Lick & 2.0 & 1800  &	2013 January 16\\
\enddata
\end{deluxetable}

\subsubsection{Abundance estimates}

We estimate abundances in the extended clouds using the model grids
from \cite{SB98}, which are computed for a typical AGN continuum shape and solar N/O ratio. They found that [N II]/H$\alpha$
is strongly abundance-dependent when either [O III]/H$\beta$ (we denote
this combination Abund1) or [O II]/[O III] (for Abund2) is used as an excitation indicator. We employ both
of these sets of line ratios (wherever our data cover the relevant
lines) on both the spectra reported here and those in Keel et al. (2012).
(In comparisons, the \citealt{SB98} ratios include both members
of the [O II], [O III], and [N II] line pairs.) Their models express the results in O/H; our 
results are listed in Table \ref{tbl-abund}. From the spectra in Table \ref{tbl-2}, we consider nuclei and
distinct clouds separately, divided where the [O III] emission shows local minima, giving up to 5
independent measurements of the EELRs in each galaxy. We include as well the abundances estimated from
line ratios in \cite{mnras2012} for nuclei and clouds, when red data covering [N II] are available
(including NGC 5252 from \citealt{Acosta1996}, leaving only SDSS 1510+07 missing), as well as comparison values
from data in \cite{cjl2009} for IC 2497 and Hanny's Voorwerp.
In deriving the line ratios, we correct for underlying H$\alpha$ and H$\beta$ absorption in the stellar
popuations using the modeled equivalent widths from \cite{Keel83} as functions of continuum color.

\begin{deluxetable}{lccccccccc}
\tabletypesize{\scriptsize}
\tablecaption{Emission-line ratios\tablenotemark{a} and abundance estimates\tablenotemark{b} \label{tbl-abund}}
\tablewidth{0pt}
\tablehead{
\colhead{Galaxy} &	\colhead{Location} & \colhead{[O III] $\lambda 5007$ flux\tablenotemark{c}} & \colhead{[O II]/[O III]} & \colhead{[O III]/H$\beta$}  &
\colhead{[N II]/H$\alpha$} & \colhead{[S II]/H$\alpha$} & \colhead{[O I]/H$\alpha$} & \colhead{Abund1} & \colhead{Abund2}}
\startdata

IC 2497		 & Nucleus &	21 & 2.16	 & 3.6 & 1.53 & 0.54 & ... & 8.69 & 8.72 \cr
Lintott et al. 2009 & 	Voorwerp & 496 &	0.110	 & 13.97	& 0.22 & 0.165 & 0.028 & 8.51 & 8.68\cr	

\cr
Mkn 1498 & Nucleus  & 490. & 	0.05	 & 4.95 &  1.33	& 0.22 & 0.07 & 8.67 & 9.30 \cr
2012 data & 	cloud & 34.	& 0.15	& 16.35 & 0.19 & 0.20	& 0.06& 8.57	& 8.56	\cr
	\cr
NGC 5252 & Nucleus & ... & 0.43 & 7.39 & 1.22 & 0.89 & 0.66 & 8.71 &  8.80  \cr
Acosta-Pulido et al. & 2-5\arcsec\ SSE & ... & 0.20 & 12.9 & 0.89 & 0.66 & 0.16 & 8.70  & 8.90 \cr
(1996)                        & 6-11\arcsec\ SSE & ... & 0.24 & 14.00 & 0.37 & 0.44 & 0.12 & 8.51  & 8.63 \cr
                   & 12-17\arcsec\ SSE & ... & 0.26 & 14.61 & 0.67 & 0.40 & 0.12 & 8.66  & 8.77 \cr
                   & 5-13\arcsec\ NNW & ... & 0.35 & 12.67 & 0.87 & 0.52 & 0.17 & 8.73 & 8.78 \cr 
		& 17-26\arcsec\ NNW & ... & 0.29 & 4.92 & 0.54 & 0.51 & ... &  8.47 & 8.71 \cr
                    \cr
NGC 5972 & Nucleus & 45.6	& ...  & 10.2 &  0.94  & 0.60	 & 0.075	& 8.69 & ... \cr		
BTA		& 8-22\arcsec\  N & 88.5	 & ... &  10.87  & 	1.01	& 0.59 & 0.119	& 8.69 & ... \cr	
		& 46-55\arcsec\  N & 3.0      & ... & 	 13.38  & 	0.77	& ...	& ... &   8.70 & ... \cr
		& 2-11\arcsec\  S &  173 & 	... & 9.28	  & 0.99 & 0.55 & 0.095	& 8.66 & ... \cr
		& 16-22\arcsec\ S &  33.8 & .. &  12.77   & 0.79	& 0.47 & 0.052	& 8.71 & ... \cr		
		& 38-54\arcsec\  S &  6.31	& ... & 22.5  &	 0.95 & 	0.95	& ... &  8.85: & ... \cr
    \cr
NGC 5972 & Nucleus &	190.& 	0.22	 & 3.40 & 0.92	& 0.49	& 0.07 & 8.57	& 8.90 \cr	
2012 data& 	cloud& 95.& 0.47	& 11.3	&  1.04& 0.61	& 0.12 & 8.72 & 8.77	 \cr
     \cr
SDSS 2201 & Nucleus &	28.2 &	2.43	& 2.67  & 1.34	& 0.83 &0.133	& 8.67 & 8.68:	\cr
BTA		& 14-20\arcsec\  S & 170 & 0.31	& 9.04  & 0.42 & 	0.27 & 0.049 & 8.52 & 8.62 \cr
		& 10-19\arcsec\  N & 186 & 0.27 & 12.22 & 0.38& 0.26 & 0.048 & 8.52 & 8.60 \cr
   \cr
SDSS 2201 & Nucleus & 	17. &	1.24	 & 7.28	& 	1.61 &	1.02	  & 0.17 & 8.77	& 8.62 \cr
2012 data & 	cloud& 78. & 0.16	& 13.4    & 	0.53 & 	0.35		& 0.07& 8.59& 8.80	\cr
      \cr
Teacup	& Nucleus &	1300.	 & 0.23 & 10.57 & 	0.31& 	0.15	 & 0.15& 	8.44& 	8.57	\cr
2012 data & 	cloud& 210. & 0.17	 & 10.73	&	0.45 &	0.36	&	0.11	& 8.52 & 8.73	\cr
	\cr
UGC 7342 & Nucleus & 	169. &  0.23 & 10.85	  & 0.73 & 0.62 & 0.083 & 8.63 & 8.82 \cr
BTA		& 8-16\arcsec\  NW & 50.0 & 0.174	 & 	14.45   & 0.26 & 0.27 & 0.055	& 8.54 & 8.58	\cr
		& 18-26\arcsec\  NW & 32.9 & 0.22 & 15.91  & 0.45 & 0.31 & 0.108 & 8.68 & 8.68 \cr
		& 11-21\arcsec\  SE  &  47.6 & 0.157 & 12.30  & 0.25 & 0.29 & 0.039 & 8.48 & 8.58 \cr
		& 23-30\arcsec\  SE & 26.9 & 0.173	 & 13.6   & 0.26 & 0.31 & 0.069 & 8.48 & 8.60	\cr
   \cr
UGC 7342 & E filament	 & 66.0 & 0.20	& 14.39  & 	0.36	& 0.34 & 0.084	& 8.56 & 8.64 \cr
    \cr
UGC 7342      & Nucleus &    140.  &  0.73  &  10.95  &  0.73 &   0.69   &   0.085 &  8.64 &   8.63	\cr
Lick PA154      &3-10\arcsec\   NW &  64.1   & ...    & 15.54     &  0.29 &   0.34  &     0.070 &  8.60    & ... \cr
               &10-23\arcsec\  NW &  35.1 &   0.37   &  18.09     &   0.60 &   0.44   &  ... &  8.78  &  8.61	\cr
                &8-14\arcsec\   SE &  22.0 &   0.51  &   20.17   &   0.46 &   0.73  & ... &   8.78 &   8.54 \cr
                &15-22\arcsec\   SE &  12.3   & ... &   16.04   & 0.24   & ... & ... &     8.48   & ... \cr
     \cr
UGC 7342   &   Nucleus &    617. &   0.15   &  10.23   &   0.72  &  0.70  &  0.084  & 8.61 &   8.82 \cr
Lick PA108    & 17-25\arcsec\ W & 57.8  &  0.15    & 9.51     &  0.43  &  0.23   & ... &  8.50 &   8.57 \cr
               & 17-33\arcsec\  E &  216.   & .. &   11.09  &  0.46 &  0.45   &   0.104  & 8.56   & ... \cr
               \cr
 UGC 11185  & Nucleus & 	1370 & 	0.20	& 9.32  & 1.66 & 	0.80& 0.170 & 8.78	& 9.10 \cr
BTA		& 4-7\arcsec\  E & 212	& 0.40	& 9.56 & 1.45	& 0.75 & ... & 	8.77	& 9.05 \cr
		& 5-14\arcsec\ W& 27.0	& ...& 5.95	 & 1.41 & 0.83 & 0.115	& 8.72 & ... \cr
     \cr
             
UGC 11185	 & Nucleus &	380.	& 0.28	& 11.82 &	1.71	& 0.88	 & 0.19	& 8.89 & 9.10 \cr
2012 data & 	cloud& 51. & 0.19	& 10.78  &	1.46	& 0.72 & 0.07	& 8.80& 9.10	\cr
       \cr
\enddata
\tablenotetext{a}{ Ratios involving [O II] use sums of the $\lambda \lambda 3726, 3729$ lines. Ratios using [O III]
give the sum of the $\lambda \lambda 4959, 5007$ lines, and those with [N II] sum the $\lambda \lambda 6548, 6583$
lens for consistency with the calculations from \cite{SB98}.}
\tablenotetext{b}{Abundance estimates are 12+log (O/H), following the model grids presented by \cite{SB98} using
[O III]/H$\beta$ versus [N II]/H$\alpha$ (Abund1) and [O II]/[O III] versus [N II]/H$\alpha$.}
\tablenotetext{c} {[O III] $\lambda 5007$ flux  within the slit from the indicated region is given in units of $10^{-16}$ erg cm$^{-2}$ s$^{-1}$.}

\end{deluxetable}

As might be expected from the broader density range present in the nuclei, the two abundance
indicators agree more closely for the low-density clouds than in the nuclei. The scatter between these values for individual locations may
be described by $\sigma = 0.19$ dex for the nuclei and $\sigma = 0.13$ dex for the extended clouds, with a systematic offset such that
values using Abund2 are larger by 0.16 dex for nuclei and 0.09 for the clouds. We thus compare clouds and nuclei for each galaxy
using single methods. 

In interpreting the results, the internal consistency allows firmer statements about relative abundances than comparison with
the solar value, in view of the uncertainties in the solar oxygen abundance. Using the solar value $ 12 + \log [O/H] = 8.76 \pm  0.05$
following \cite{Caffau2008}, most systems have clouds with subsolar metallicity, and the
clouds are systematically lower in O/H than the nuclei. The mean values are $ 12 + \log [O/H] = 8.62$ from Abund1 ($\sigma = 0.11$)
and 8.70 ($\sigma=0.15$) from Abund2 for the extended clouds, versus 8.67, $\sigma =0.11$ and 8.84, $\sigma=0.22$ for the nuclei.
(The samples are not identical for the two abundance indicators since we do not always have [O II] data.) On this scale, several of the
galaxies have extended clouds with abundances as low as 0.5 solar. In such cases as NGC 5252, where the data include multiple abundance estimates
and we see systematic changes with radius, we may speculate about mixtures of nuclear outflow and more pristine, recently acquired gas in tidal features.

The uncertainty in these abundance estimates is almost certainly dominated by such details as ranges of density, which introduce ranges
of ionization parameter and hence line ratios beyond the model assumptions. Within the low-density extended clouds, each indicator is internally consistent; for galaxies
where our spectra sample distinct multiple emission stuctures (often on opposite sides of the galaxies), the scatter about the mean for each
system is $\sigma = 0.07$ for Abund1 and $\sigma=0.04$ for Abund2. 
At this level of confidence, these results show that some systems are
significantly subsolar throughout (SDSS 2201+11, UGC 7342,
IC 2497/Hanny's Voorwerp) while others are closer to solar
abundances (Mkn 1498, NGC 5972, the Teacup, UGC 11185). This distinction
may reflect different contributions from gas originally associated with
the host galaxies versus external sources such as tidal disruption of
low-mass, low-metallicity galaxies.

\subsection{Velocity mapping}

For NGC 5972, SDSS 2201+11, UGC 7342, and UGC 11185, we obtained full velocity fields from \oiii 
emission using the 
SCORPIO-2 system at the 6-meter Large Altazimuth Telescope (BTA) in its 
scanning Fabry-Perot interferometer (FPI) mode. The choice of targets was set both by weather and 
the availability of appropriate blocking filters to
isolate \oiii $\lambda 5007$ at the galaxy redshift. (UGC 7342 was observed in the $\lambda 4959$ line to use an available filter). 
In addition, we re-reduced the similar \oiii data for NGC 5252 originally presented by \cite{Moiseev2002},
\cite{Afanasiev2007a}, and \cite{Afanasiev2007b}.
 The observational parameters are listed in Table \ref{tbl-btamaps}.

Data reduction followed the procedures in
\cite{MoiseevEgorov}. Voigt-profile fits of the data cube as sliced in the spectral direction were used to generate spatial
distributions of continuum intensity, continuum-subtracted line intensity, mean radial velocity,  and velocity dispersion.
The Fabry-Perot results are summarized graphically in Fig. \ref{fig-btamaps}, with the HST \oiii images aligned for comparison.
In UGC 11185, where the long-slit data show multiple \oiii emission components, we have examined the data cube to investigate its
two-dimensional extent on the sky; the component to the west of the nucleus at $cz  \approx 12700$ km s$^{-1}$ is associated with the
roughly conical structure just east of the nucleus in the HST images. 

\cite{Harrison} have recently presented integral-field spectroscopy of the Teacup system, giving velocity mapping of the entire
emission region. They show that the large emission-line loop coincides with a radio-continuum structure, and identify a
kpc-scale outflow region near the AGN itself. We incorporate these results in our later discussion.

\begin{figure*} 
\includegraphics[width=150.mm,angle=0]{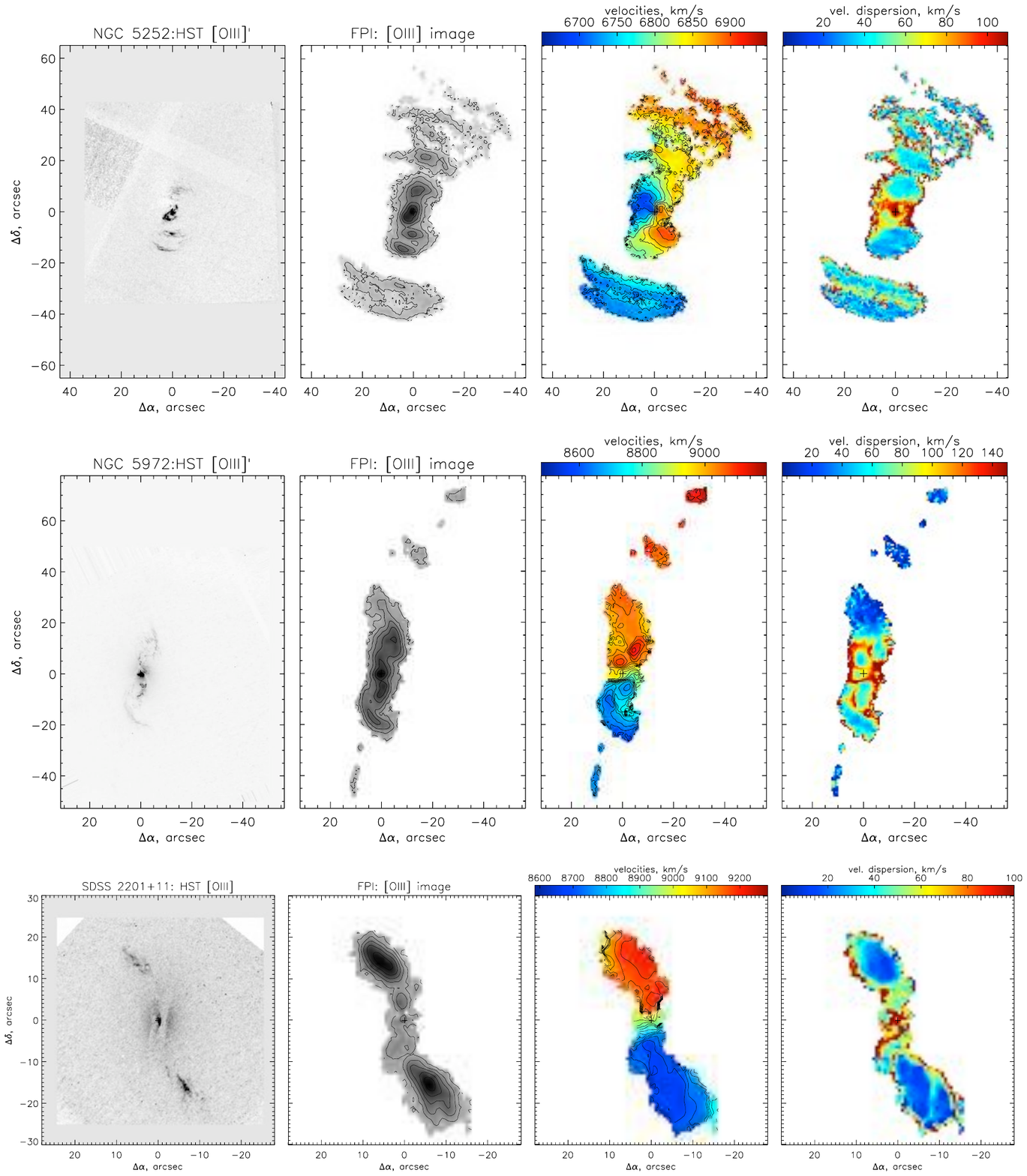} 
\caption{Summary of BTA Fabry-Perot observations, showing line images from the HST images and as derived from
the Fabry-Perot data cubes, radial-velocity centroid and 
velocity dispersion from Voigt profile fitting, corrected for the instrumental resolution in each case. In addition to the 
centroid velocities shown here, NGC 5972 has regions with asymmetric or split line profiles to the north and south of the nucleus. The western
component of UGC 11185 falls outside the monochromatic field of the ACS ramp fiters and is thus not well shown in the HST \oiii image.
(Continued on next page)} 
\label{fig-btamaps}
\end{figure*}

\addtocounter{figure}{-1}

\begin{figure*} 
\includegraphics[width=140.mm,angle=00]{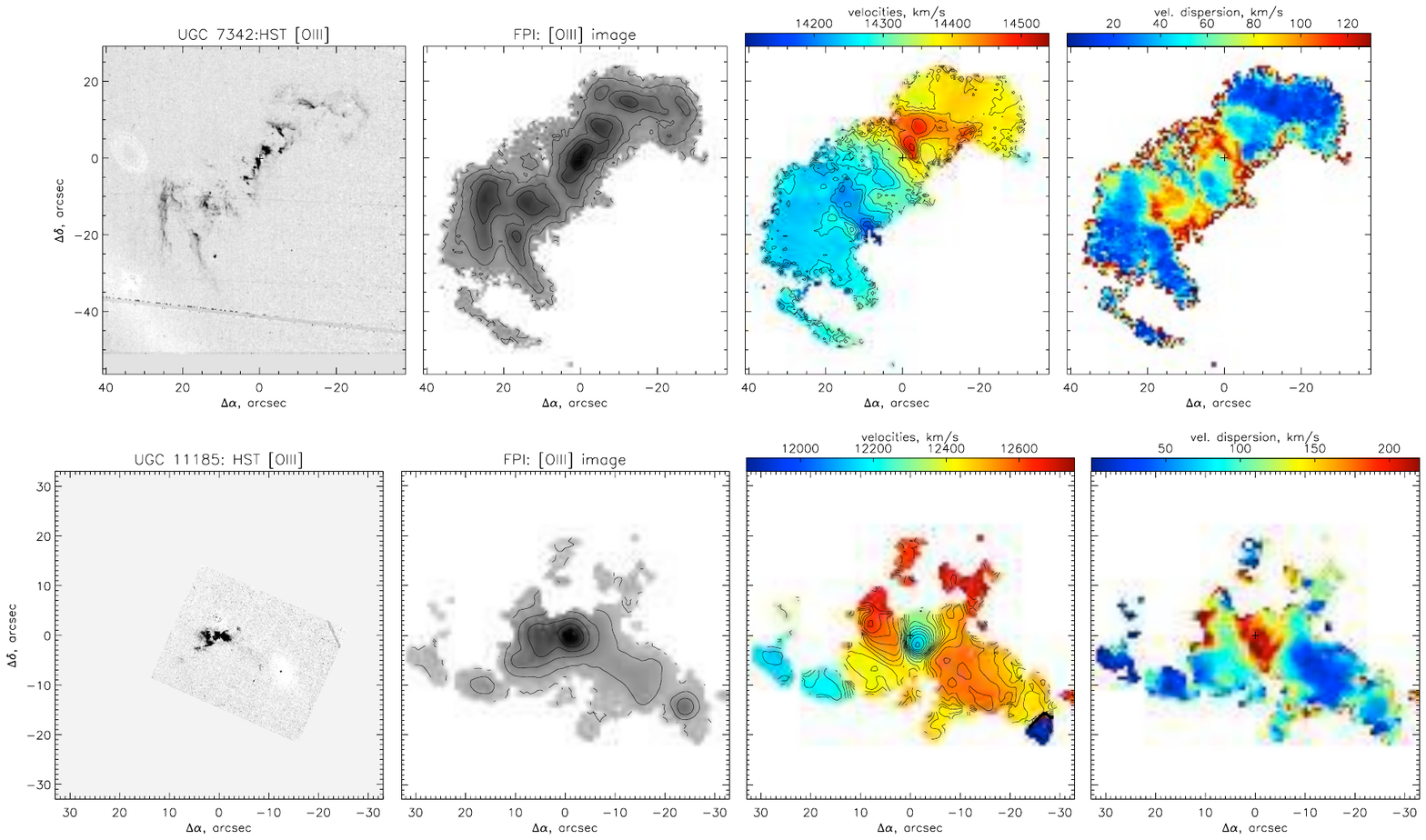} 
\caption{(continued)} 

\end{figure*} 

\begin{deluxetable}{lcccl}
\tablecaption{Fabry-Perot velocity maps \label{tbl-btamaps}}
\tablewidth{0pt}
\tablehead{
\colhead{Galaxy} &	\colhead{Spectral FWHM[km s$^{-1}$]} & \colhead{Seeing FWHM [\arcsec]} & \colhead{Exposure times [s]}  &  UT Date }
\startdata
NGC 5252		&	67		&	2.2		&	$450 \times 24 $ & 	1998 February 24/25     \cr
NGC 5972		&	140		&	2.2		&	$180 \times 30$ &  2014 May 21/22 \cr
SDSS 2201+11	&       42 		&      2.0 		&       $300 \times 30$ &   2012 August  23/24  \cr
UGC 7342		&	137		&	2.8             &       $ 240 \times 30$ &    2014 April 0/1 \cr
UGC 11185                &     140            &      2.7             &       $ 210 \times 30$     &    2014 October  20/21  \cr
\enddata
\end{deluxetable}

\subsection{Ground-based imaging}

The SCORPIO--2 system at the BTA was also used in its direct--imaging mode for 
$V$-- and $R$--band images of SDSS 2201+11 and UGC 11185. At a pixel scale 0.357\arcsec/pixel covering a field 6.1\arcmin square, these have excellent 
surface-brightness sensitivity, tracing tidal features beyond the sensitivity of the HST intermediate-band data. Especially for
SDSS 2201+11, the V images also show faint extended parts of the \oiii clouds.

For NGC 5972, where both the emission-line regions and stellar tidal tails extend beyond the HST field, we supplement those data with images from the 
KPNO 2.1m telescope, using a $V$ filter for the continuum and a filter nominally centered at 5142 {\AA}  with FWHM 174 {\AA}, for redshifted \oiii . A 
TI CCD gave 0.30{\arcsec} pixels. Total exposure times were 90 minutes in the \oiii filter and 30 minutes in the $V$ continuum.

To seek tidal features around SDSS 1510+07 beyond the SDSS surface-brightness limit, we initially collected a total exposure of 80 minutes in $R$ with the SARA remote
0.6m telescope at Cerro Tololo, and followed up this detection using  a total exposure of 5 hours in two sessions with the
SARA 1-m telescope at Kitt Peak. Surface-brightness calibration used standard stars from \cite{landolt}.

As part of our examination of the interaction history of the EELR objects without energy-budget shortfalls, we use archival and new BTA images of Mkn 78
in the continuum (filters SED607 and FN641). The new FN641 exposure was obtained in October 2014 to test the possibility of faint reflections in the
rich stellar field toward Mkn 78.

\section{Host-galaxy structure}

\begin{figure*} 
\includegraphics[width=170.mm,angle=0]{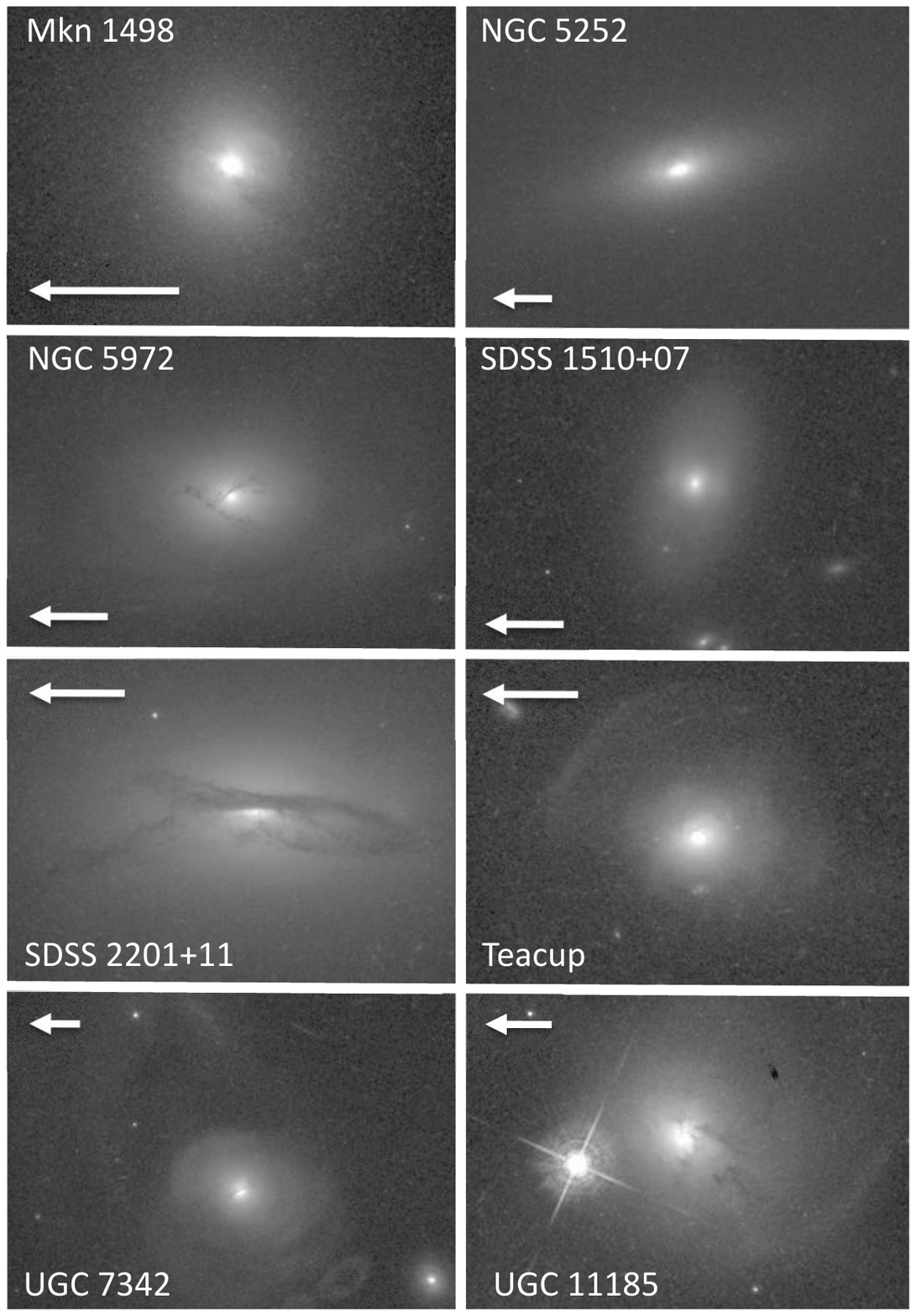} 
\caption{Host galaxy images in the HST F621M filter, shown in an offset log scaling, with
brightness and contrast independently scaled for each. 
NGC 5252 is shown from a WFPC2 image in the F588M band. The Mkn 1498 and Teacup images have been corrected for emission-line contamination. 
Each image is shown with north to the left and east at the bottom; as in Fig. \ref{fig-overview1}, arrows are 5{\arcsec}  long and point north.} 
\label{fig-montage} 
\end{figure*}

Fig. \ref{fig-montage} shows the HST continuum images of the
galaxies, displayed to span a wide range in surface brightness.
Several show obvious tidal structures in these images, and all
show evidence for disturbance via tidal tails or disrupted dust structure. We discuss thes ein turn, and compare to the analogous
set of AGN with EELRs but energetics consistent with obscuration rather than fading.

\subsection{Merger and interaction signatures}

{\bf Mkn 1498} shows rich structure not resolved in earlier images. There are
two broad spiral bands, which persist in residuals after both
ellipse fitting (\S \ref{s-surfacephotometry}) and using a model with fixed orientation (which is not influenced by these features). These may be the aftermath of a merging
event. 

{\bf NGC 5252} shows a symmetric, edge-on stellar disk in a luminous bulge. The evidence for past interaction here consists of the
small-scale half-spiral of dust seen near the nucleus, and the kinematically decoupled stars and gas \citep{morse}.

\begin{figure} 
\includegraphics[width=70mm,angle=00]{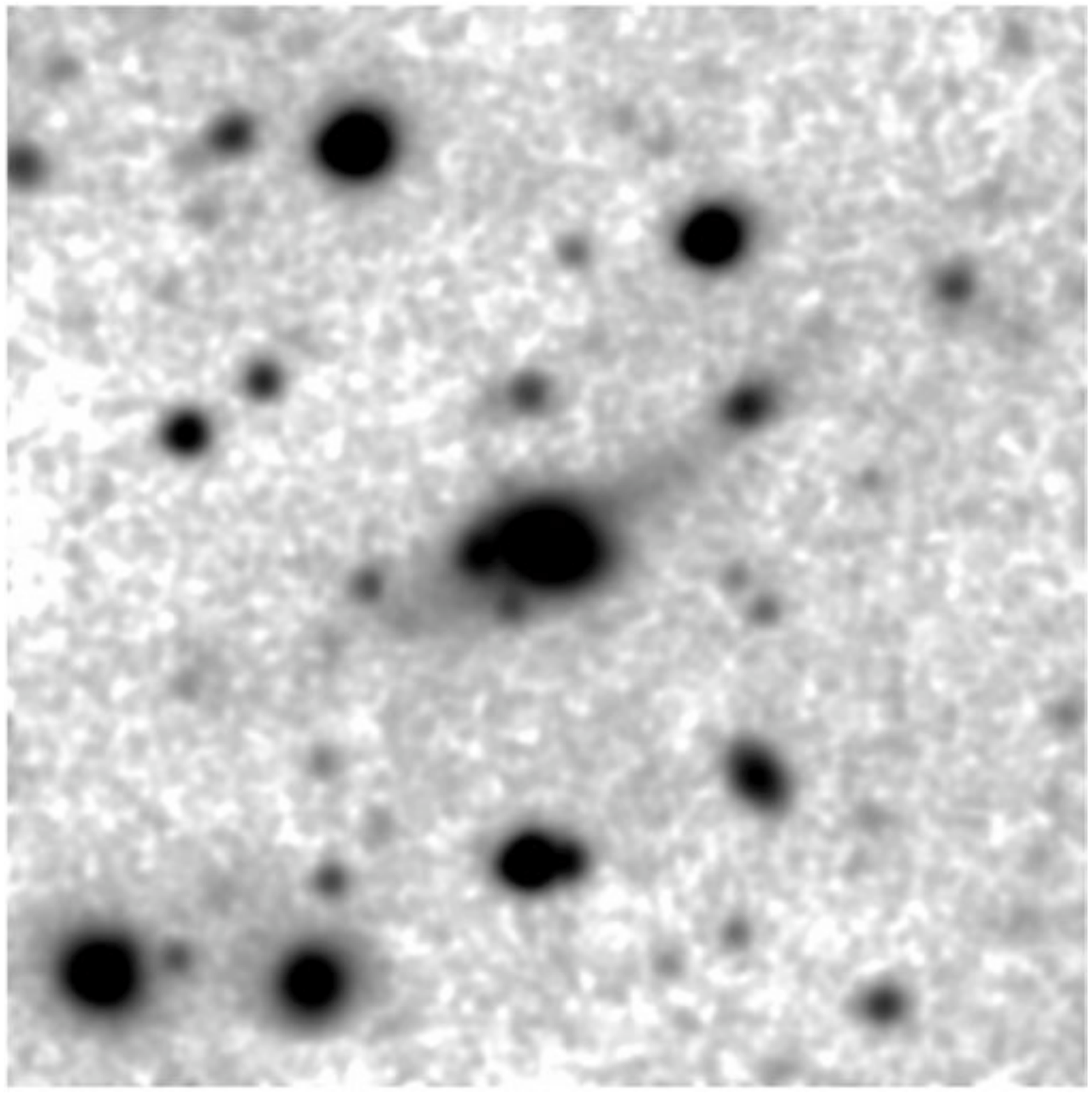} 
\caption{Deep $R$ image of SDSS 1510+07 from the SARA 1m telescope, smoothed by a Gaussian of
FWHM=2.3\arcsec. A tail connecting to the companion galaxy NW is detected here, and a more diffuse asymmetric
extension on the SE side
of the AGN host galaxy. The field is 199{\arcsec} square with north at the top, with an offset log intensity mapping.} 
\label{fig-sdss1510sara}
\end{figure}

{\bf SDSS 1510+07} has a small dust patch within the bar, showing 
the south to be the near side. This is detected in both continuum bands 
from independent fitting. This dust lane is closest to the nucleus on its western end. There is some evidence in the F621M image for
a symmetric counterpart on the far side, with its deepest pixels showing
a flux deficit $\approx 10$\% rather than $\approx 28$\% on the near side.
Our ground-based images show a tail connecting the AGN host galaxy to another galaxy
just to the NW, SDSS J151001.83+074058.4, at typical surface brightness R=25.7 mag arcsec$^{-2}$,
as well as more diffuse asymmetric structure outside the disk to the southeast (Fig. \ref{fig-sdss1510sara}). This tail is at roughly right angles to the
axes of the ionized clouds.

\begin{figure} 
\includegraphics[width=70mm,angle=00]{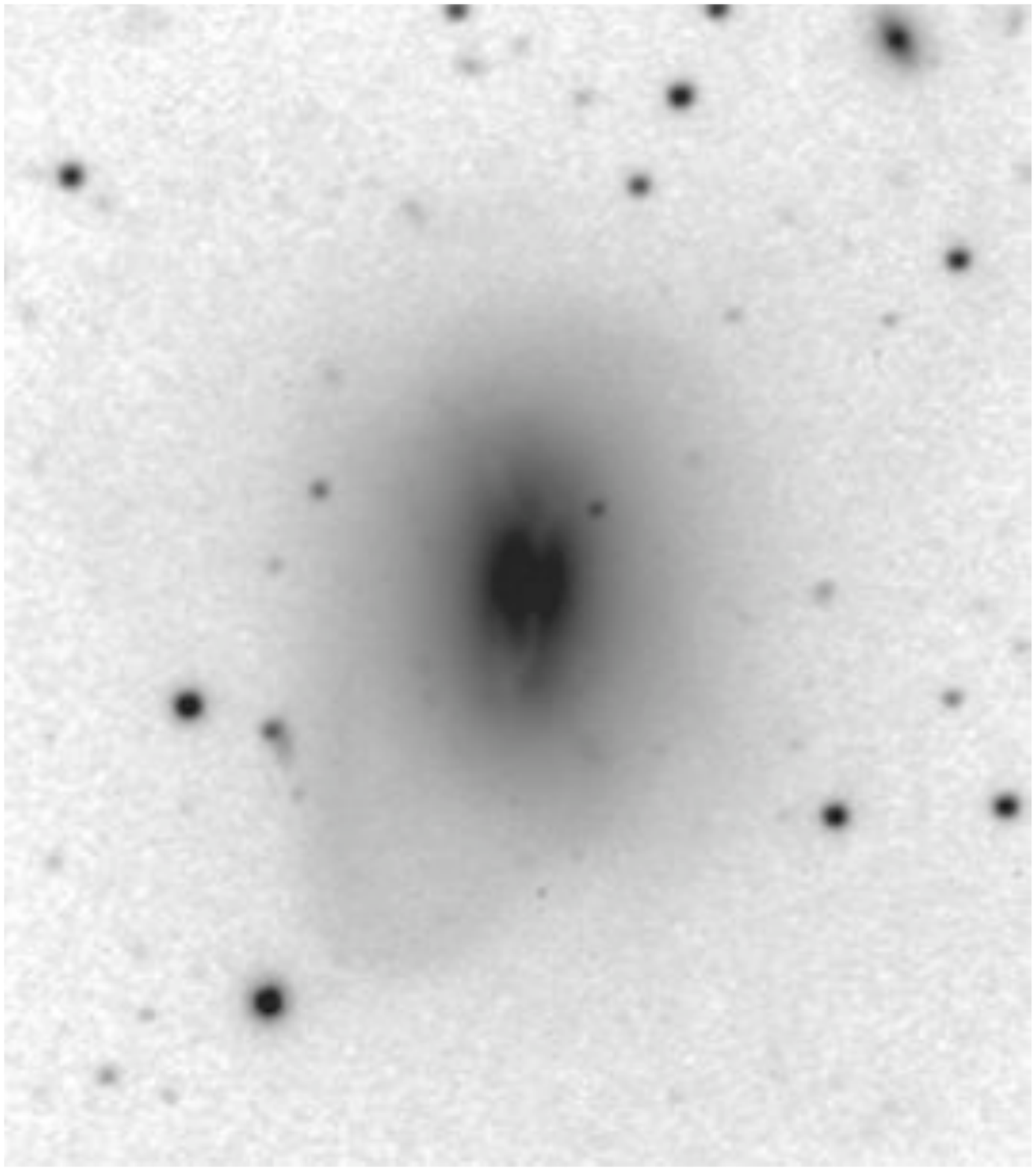} 
\caption{BTA $R$-band image of SDSS 2201+11 from a 5-minute exposure, showing outer tidal features.
The emission clouds show weakly via the H$\alpha$+\nii  lines.The field shown spans 96{\arcsec} $ \times$ 109{\arcsec} ,
 with north at the top, displayed using an offset logarithmic intensity mapping.} 
\label{fig-btasdss2201r}
\end{figure}

Beyond the twisted dust lanes apparent in the HST data, the stellar envelope of {\bf SDSS 2201+11} shows asymmetry and outer loops indicating
an aging merger (Fig. \ref{fig-btasdss2201r}).

The {\bf Teacup} host galaxy clearly has asymmetric structure, well described as 
a tidal tail with a shell-like feature at radius 10.0{\arcsec} on the opposite 
side. These structures resemble stages in the merger of a small, 
dynamically cold system with a bulge-dominated one, as in modeling of shell 
galaxies by  \cite{hernquinn}. This configuration
is seen in their models of radial encounters, early in the postmerger phase when only 
a single shell appears. The tidal tail distinguishes this origin for
the shell or arc from the weak-interaction case modeled by \cite{thomson}. Generally, \cite{feldmann}
find that such narrow tidal features around elliptical galaxies must result from
acquisition of cold disk systems with mass ratios $\approx 0.1$. and may remain distinct for 1-2 Gyr.

In interpreting the tails of {\bf UGC 7342}, the merger is more
advanced than it first appears; our spectra show that the apparent 
companion to the southeast (SDSS J121820.18+291447.7, seen at the right
edge of the frame in Fig. \ref{fig-overview1})
is a background star-forming system at $z=0.0674$, far beyond
UGC 7342 at $z=0.0477$; it shows positive residuals in our \oiii continuum-subtracted image
only because its H$\beta$ and \oiii $\lambda 4959$ lines are in opposite wings of the
ramp-filter transmission at this setting. This identification
reduces the computed far-IR luminosity of UGC 7342, since the
WISE images \citep{WISE} show that the background system 
strongly dominates the flux
at 22$\mu$m. The IRAS and {\it Akari} limits for the blended
combination \citep{mnras2012}
must be reduced to account for this unrelated contribution. Tidal tails
are seen to 37{\arcsec} (35 kpc), with several distinct shell-like surface 
brightness steps around 10{\arcsec} radius (9.5 kpc).

All our faded-AGN candidates are found in interacting,
merging, or postmerger systems. Since the fading timescales
we derive are of the order of the light-travel times across the systems,
in contrast to the dynamical timescales which are orders of magnitude 
longer, we would expect the same to be true of the non-fading (that is,
mostly obscured) systems in the parent sample of AGN with large-scale
ionized clouds from \cite{mnras2012}. 
Of the 12 obscured-AGN host galaxies in this parent sample, only one has
no obvious source for extraplanar gas (merging, postmerger, interacting, or ram-pressure
stripping in Virgo in the case of NGC 4388). This object is {\bf Mkn 78}, in an elliptical host system. 

This prompts us to re-examine Mkn 78 to seek potential subtle tidal features.
Using pre-refurbishment HST
images, \cite{whittlewilson} report a dust lane crossing the inner 
$\approx 1${\arcsec} at a skew angle to the major axis, which they suggest may
result from an advanced accretion event, one old enough to leave
little or no trace in the outer isophotes.
For more insight into the structure of Mkn 78, we analyze the BTA images
in BVR and red-continuum bands from \cite{smirnova2010}, and a new exposure in the
FN641 filter centered near 6410 \AA\ .
The deep image in the SED607 filter, which lies between \oiii and [O~{\sc{i}}] $\lambda 6300$
lines, is especially important, rejecting the bright emission-line 
clouds. As Smirnova et al. note, the images show no asymmetric 
structure or shells to deep levels, to surface brightness
$R \approx 26$ mag arcsec$^{-2}$.  A resolved companion or tidal plume
appears in both these data sets and, in hindsight, the SDSS images, about 12{\arcsec}  W of the nucleus. We
isolate it by fitting elliptical isophotes to the galaxy after masking this region and contaminating objects,
and examine the residuals after subtracting a model based on these isophotes. These also reveal
an asymmetric feature oriented E--W, which could be interpreted as a small-scale stellar disk seen edge-on,
and a light deficit crossing the nucleus near PA 35$^\circ$,
as well as the diffuse excess. This excess light, detected at 8--16{\arcsec}  from the core, is asymmetric with a tail to the north, and involves roughly
$6 \times 10^{-4}$ of the galaxy's red continuum light. A tidal origin is plausible, although not fully demonstrated; this would mean that all EELRs
in our parent sample are in interacting, merging, or ram-pressure stripped systems.

\subsection{Surface-brightness profiles}

Photometric profiles in most of these galaxies (derived as in \S \ref{s-surfacephotometry})
show them to be strongly bulge-dominated, in some cases with
no significant disk component. The averaged brightness profile in the Teacup galaxy
is close to a de Vaucouleurs $r^{1/4}$ form over nearly 10
magnitudes in surface brightness and a range 3-370 pixels (0.15--18\arcsec) in radius.

To quantify the bulge properties, and derive detailed information or limits on disk components, we used the
GALFIT software (\citealt{peng2002}, \citealt{peng2010}), version 3.0.5. Dust masks were produced in a semi-automated way, starting with 
the dust maps in Fig \ref{fig-dustmontage}, using
attenuation thresholds depending on the local signal-to-noise ratio and manually selecting the threshold appropriate to various parts of the
image. Interloper star and galaxy images were simply masked from the fits. Central PSF components to accommodate an unresolved AGN
were allowed, but significant only in Mkn 1498, NGC 5252, and the Teacup AGN.
For galaxies observed with WFC3, the F763M image was fit, to
reduce the effects of dust and young stellar populations. For NGC 5252, we fit the F606W WFPC2 image.
The Sersic indices of ``bulge" components from these fits are listed in Table \ref{tbl-galfit}. Here, $n$ is the single-component 
Sersic index, and $n_d$ is the index when an additional component fixed to be an exponential disk is included. Disk/bulge and AGN/bulge ratios
are in intensity. All $n$  values
fall into the domain occupied by classical bulges, rather than pseudobulges or other structures \citep{FisherDrory}. 

\begin{deluxetable}{lcccc}
\tablecaption{Fitted Sersic indices \label{tbl-galfit}}
\tablewidth{0pt}
\tablehead{
\colhead{Galaxy} &	\colhead{$n$} & \colhead{$n_d$} & \colhead{Disk/bulge} & \colhead{AGN/bulge }}
\startdata
Mkn 1498 &  6.04 & 6.14 & 0.41 & 0.06 \\
NGC 5252 &  3.55 & 5.18 & 0.13 & 0.01 \\
NGC 5972 &  4.26 & 3.71 & 0.51 & ... \\
SDSS 1510+07  &  4.6 & 2.41 & 1.58 & ... \\
SDSS 2201+11 &  3.51 & 7.12 & 0.44 & ... \\ 
Teacup &  5.94 & 5.65 & 0.25 & 0.02 \\
UGC 7342 &  3.95 & 4.61 & 0.09 & ... \\
UGC 11185 &  7.31 & 8.87 & 0.02 & ...\\
\enddata
\end{deluxetable}

In SDSS 1510+07, the stellar disk was also retrieved by GALFIT, although simultaneous fits to the bar and disk are poor with 
this set of components. In some of the highly disturbed systems, the physical significance of a formal exponential-disk fit is
not clear. 

Some potential explanations for the fading-AGN phenomenon involve interactions between members of binary black holes; several sets of
simulations show the individual accretion disks having rapidly fluctuating luminosities (e.g., \citealt{bogdan}). One signature of
such a process would be ``scouring" of stars in the core (\citealt{HopkinsHernquist}, \citealt{Rusli}). In this sample, appropriately accurate measurements
of the innermost light profile proved infeasible because of dust lanes or excess light from the AGN itself.

\subsection{Dust structures}

Dust can show whether the cool ISM in a galaxy is relaxed, and since
its attenuation effects strongly favor detection on the near side of a 
galaxy, provide some three-dimensional information on its location.
We examine dust attenuation primarily through the ratio of
continuum images to smooth elliptical models generated from the
surface-brightness profiles (\S \ref{s-surfacephotometry})). This procedure may produce
artifacts, especially near the cores or dust lanes, so we also compare the 
results to ratio maps between the two continuum wavelengths. These color 
maps have lower signal-to-noise ratios for detection, since the wavelength
baseline is small, but do not suffer from the modeling uncertainties
that can occur with galaxy intensity.

\begin{figure*} 
\includegraphics[width=135.mm,angle=270]{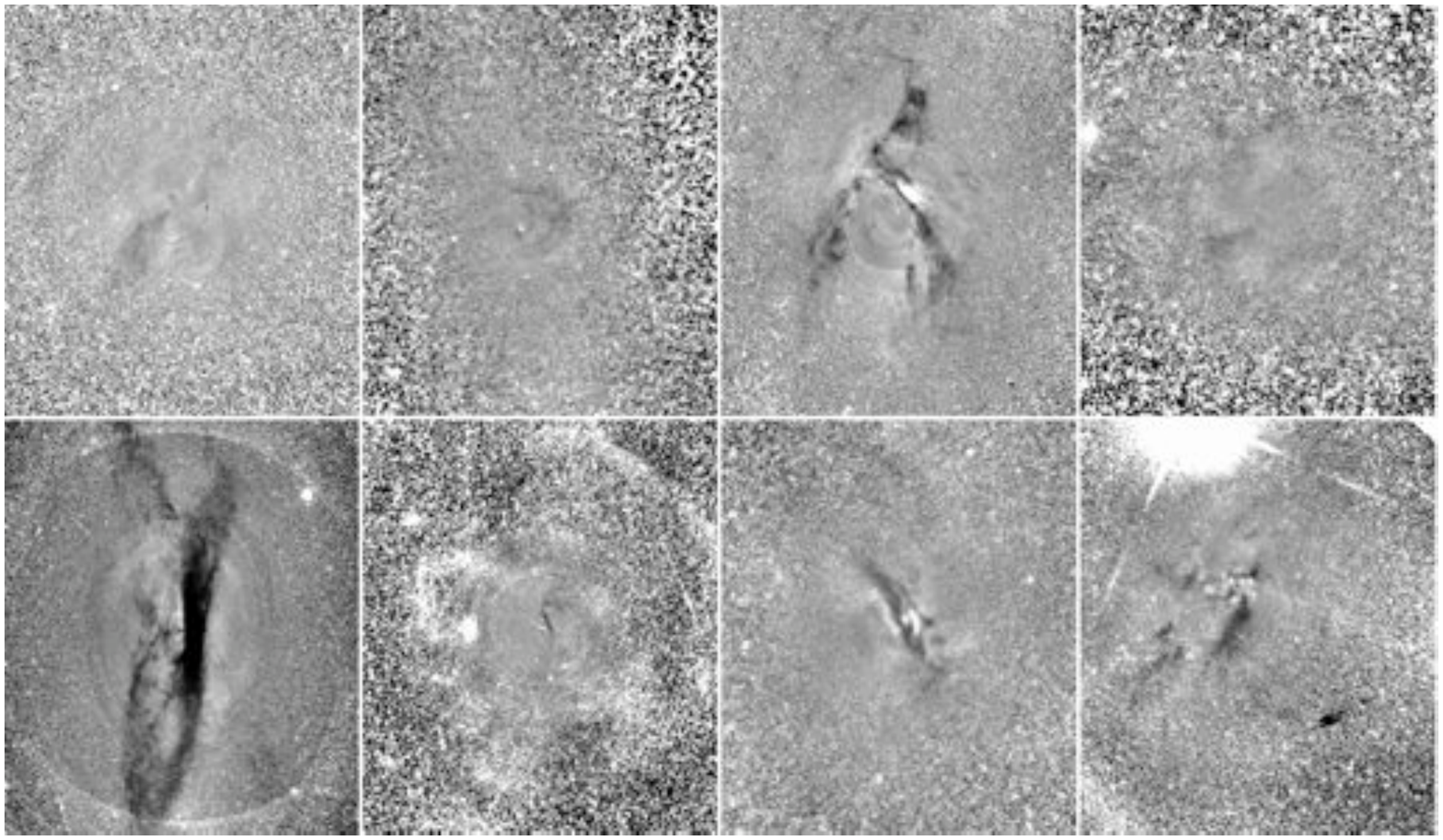} 
\caption{Attenuation images of target systems, using a common linear intensity display running from
0--1.5 in ratio of image to model. Ordered as in Table 1, the top row includes (left to right) Mkn 1498, NGC 5252, NGC 5972, and SDSS 1510+07;
the bottom row has SDSS 2201+11, the Teacup, UGC 7342, and UGC 11185.
The field is 18.0{\arcsec}  E--W, except for SDSS 1510+07 and UGC 7342 at 9.05{\arcsec}, and UGC 11185 at 15.0{\arcsec}.
North is at the top and east to the left.}
\label{fig-dustmontage}
\end{figure*}

These attenuation maps in the bluer continuum image (WFC3 F621M or WFPC2 F588N) are shown in Fig. \ref{fig-dustmontage}, to a common scale in intensity
ratio. Complex dust lanes and patches are ubiquitous, often with
distributions suggesting the aftermath of strong interactions or
mergers. Only in SDSS 1510+07 do we fail to detect such 
disorganized dust structures.

In {\bf Mkn 1498}, we see an irregular and asymmetric 
dust lane at a sharp angle to the emission structures, detected to 5\farcs 6 (6.2 kpc) toward the southeast.

In {\bf NGC 5252}, we find a small half-spiral of dust seen to the northwest of
the nucleus, strongly misaligned with the
stellar disk, as noted by \cite{tsvetanov}. This disk has
an orientation which could be circular in the plane defined by
the kinematics of the extended ionized gas filaments \citep{morse}
and, at modest resolution, H I \citep{PrietoFreudling}.

The dust in both {\bf NGC 5972} and {\bf SDSS 2201+11} crosses the central region multiple times on opposite sides of their nuclei. In both cases,
this may represent a side view of a differentially precessing
disk \citep{SCKD1992}; such a disk could continue to the ionized clouds in SDSS 2201+11,
where a dust lane connects to the northern cloud (and in order to show distinct absorption, this feature
must lie on the near side of the galaxy). Our long-slit and Fabry-Perot data show the northern side of the system to be receding,
helping to specify the geometric conditions of our view.

In the {\bf Teacup system}, the ratio images show detail in the complex dust features seen even in the unprocessed data, especially to 
the southeast of the nucleus with a connection spiraling
outward to projected distance 1.9\arcsec (3.3 kpc). The deepest attenuation in this image, seen 0.3{\arcsec} W of the nucleus, has 0.41 of the model intensity in the green band 
and 0.58 in the red. The inner part of the dust band runs roughly north-south, at
about 45$^\circ$ to the greatest elongation of the ionized gas.

{\bf UGC 7342} shows a warped dust lane near the core, which may hide the AGN
itself from direct optical view.

\subsubsection{Precessing disks}

Several of these systems show dust lanes resembling the complex
crossing geometry in NGC 4753 \citep{HubbleAtlas}, which prompted us to consider them
in the light of the model for differentially-precessing disks of 
material developed by \cite{SCD1988} and applied
to NGC 4753 by \cite{SCKD1992}. Using their scale-free logarithmic potential model with a flat rotation curve,
a ring of gas in an orbit of radius $r$ at circular velocity $v_c$
and inclination $i$ to the principal plane of a flattened potential well, will precess at
angular rate
$$ \dot{\Omega}_p = - (3v_c / 4 r) \eta \cos i  \eqno{(1)}$$
where $\eta$ depends on the shape of a (generally triaxial) potential: for major and minor axes $a,c$,
$$\eta = {{2}\over{3}} \ {{1-(c/a)^2} \over {2+(c/a)^2}} \eqno{(2)}.$$
In applying the model, we assume that the potential shape is oblate with 
the same axial ratio as the starlight distribution (that is, that we see
the mass distribution edge-on), giving $\eta=0.113-0.128$. Since \cite{SCKD1992}
find that the damping term in $i$ corresponding to an effective viscosity is small, we ignore
this effect.

\begin{figure*} 
\includegraphics[width=130.mm,angle=270]{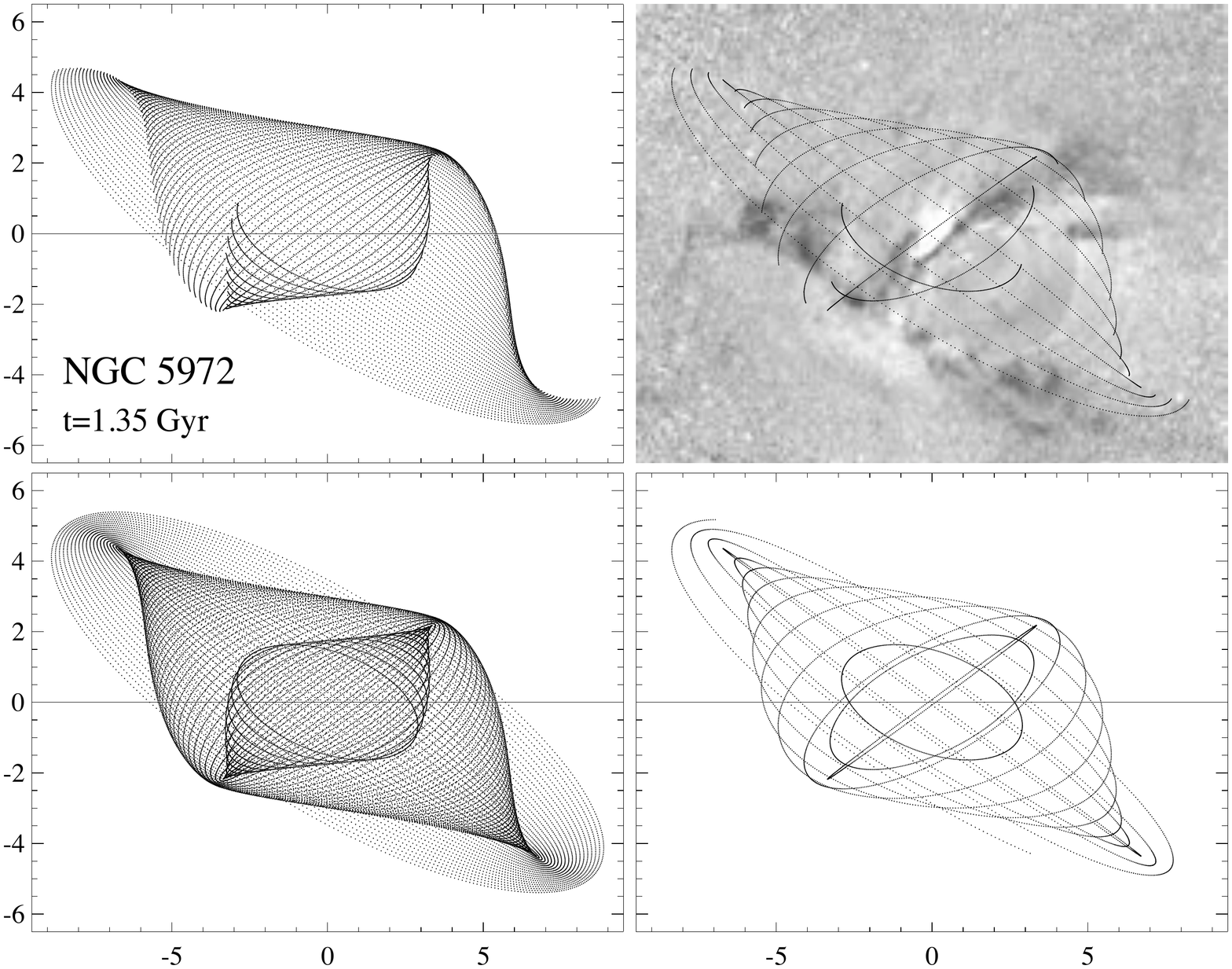} 
\caption{Schematic precessing-disk model for dust in NGC 5972. These are edge-on views at a time 1.35 Gyr after introduction of an annulus of material
inclined by $i=33^\circ$ to the principal plane of the host galaxy, a timestep which captures three important dust locations in the images. 
In comparison to images of NGC 5972, north would be to the left.
The upper panels show only the material on the front side, which might be visible through attenuation of starlight, while the lower
panels show the entire configuration. The left panels show rings of tracers spaced radially every 0.1 kpc, while the right panels have 
0.5-kpc spacing to indicate the role of dust in distinct lanes. Scales show offsets in kpc. The upper right panel, with spaced rings shown
where they cross in front of the system, is superimposed on the dust-attenuation image from Fig. \ref{fig-dustmontage}.} 
\label{fig-ngc5972disk}
\end{figure*}

We implemented their model for NGC 5972 and SDSS 2201+11 to explore the ranges of geometries and
times that could match the dust geometry we observe; these are the systems with well-resolved structure
including features expected in the model. The starting point is an idealized introduction of
a flat annulus of interstellar matter to the host galaxy at $t=0$ at an inclination to the equatorial plane
estimated from the crossing angles of dust lanes; $33^\circ$ for NGC 5972 and
$21^\circ$ for SDSS 2201+11. The circular velocity is taken from averaging our long-slit data; 190 km s$^{-1}$ for
NGC 5972 and
255 km s$^{-1}$ for SDSS 2201+11. This fixes the radial scale, leaving the age and viewing geometry as free parameters.
Seeing discrete crossing dust lanes, rather than a broad projected distribution as in Centaurus A,
would mean that the dust is confined to narrow radial zones, affording an additional freedom in matching predictions
(and making our efforts more of an existence proof than a detailed fit)
since there is little additional information on their line-of-sight locations. The model must be old enough for dust lanes
to cross at sharp angles as we observe, a limit which in turn depends on the radial distance range between the dust features.
For NGC 5972, dust is detected in our ratio images to about 9{\arcsec} or 5.5 kpc. Since dust detection is strongly favored in front of 
the bright inner bulge, we consider dust in rings from 3-10 kpc in radius. From a typical direction along the galaxy's equatorial plane, this
leads to crossing of the inner rings as soon as 0.9 Gyr, and up to 3 Gyr for the outer rings (Fig. \ref{fig-ngc5972disk}). For dust absorption, we examine
only the parts of these precessing rings on the near side, while for comparison with the ionized gas, we include the far side as well.
In NGC 5972, the inner ionized-gas regions might be interpreted as the intersection of ionization cones with such rings of material.
Supporting this idea,
there is no emission associated with the dust lanes in front of the core, with some emission regions plausibly connected with one such lane 
when projected $\approx 3$ kpc from the nucleus.

\begin{figure*} 
\includegraphics[width=130.mm,angle=270]{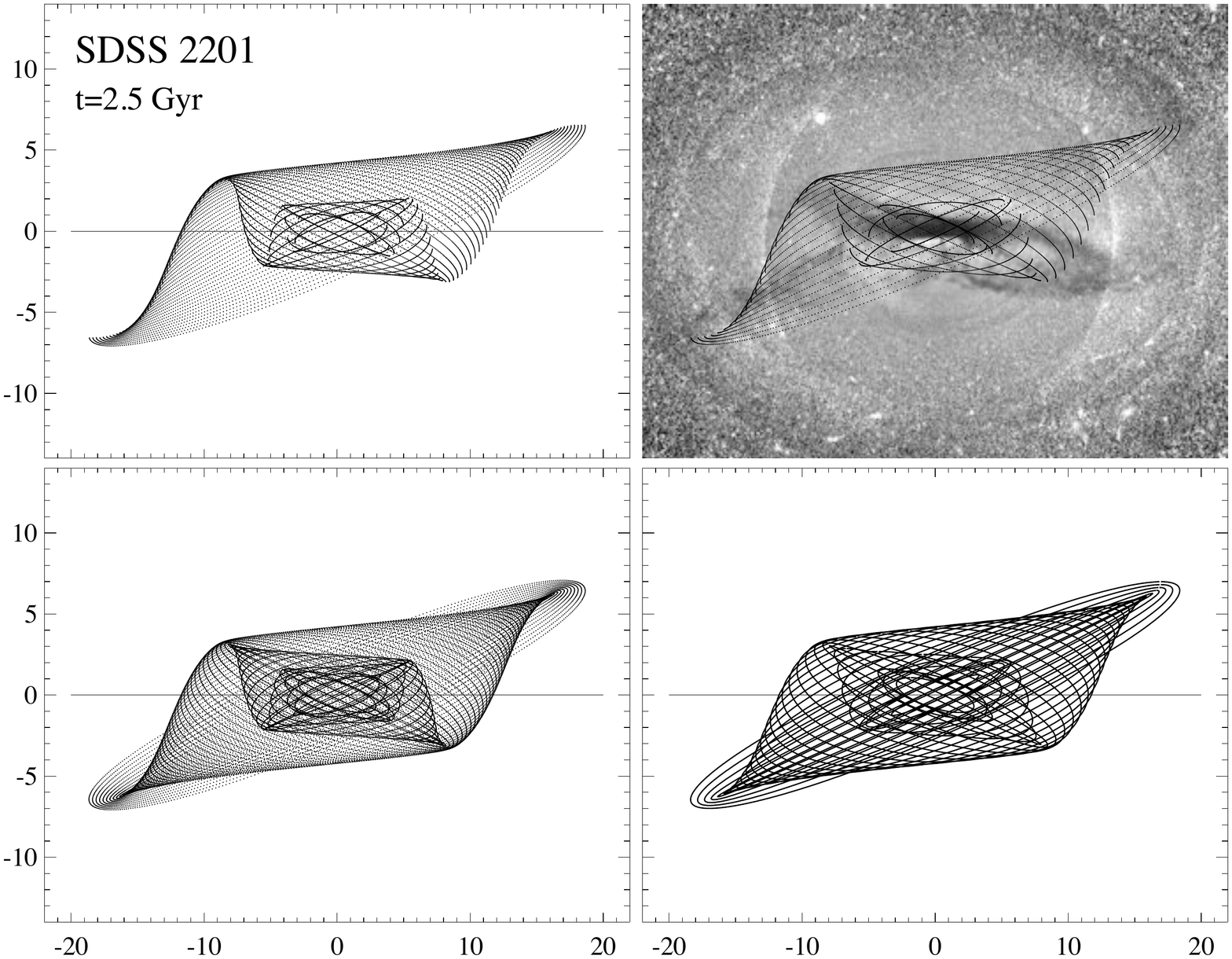} 
\caption{Schematic precessing-disk model for dust in SDSS 2201+11. These are edge-on views at a time 2.5 Gyr after introduction of an annulus of material
inclined by $i=21^\circ$ to the principal plane of the host galaxy, a timestep which captures three important dust locations in the images. 
In comparison to images of SDSS 2201+11, north would be to the left.
The upper panels show only the material on the front side, which might be visible through attenuation of starlight, while the lower
panels show the entire configuration. The left panels show rings of tracers spaced radially every 0.1 kpc, while the right panels have 
0.5-kpc spacing to indicate the role of dust in distinct lanes. Scales show offsets in kpc. The upper right panel, with spaced rings shown
where they cross in front of the system, is superimposed on the dust-attenuation image from Fig. \ref{fig-dustmontage}.} 
\label{fig-sdss2201disk}
\end{figure*}

In SDSS 2201+11, there is a direct connection  between gas and dust distributions;  a dust lane leads from the main dust disk into the
NNW emission-line cloud (which is thereby on the near side, and receding as shown by both the longslit and
Fabry--Perot observations). In modeling this system, we take an axial ratio of 0.75 for the host starlight and
inclination $i= 21^\circ$. Dust is detected to a radius 16{\arcsec} or 9.5 kpc; we consider annuli from 3-10 kpc. In this system we can evaluate the
circular velocity from the full 2D BTA velocity map, adopting 255 km s$^{-1}$. Crossing of dust lanes near the core starts around 1.4 Gyr, occurring later
for features farther out. A sample model is shown in Fig. \ref{fig-sdss2201disk}.

If the ionized gas forms part of the outermost ring matching this model, its actual cone angle for illumination from the nucleus is substantially larger than the
$23^\circ$ we see in projection, since we see the warped disk at a sharp angle.

\section{Star clusters}

The evidence for a region of outflow-triggered star formation in Hanny's Voorwerp \citep{HSTHV}, and more generally the signs of a role for AGN feedback in
regulating star formation (e.g., \citealt{Fabian}), give impetus to search for young stars and clusters associated with these AGN and surrounding gas clouds. These may be detected either 
directly, when in luminous blue clusters, or secondhand, as localized regions with \oiii/H$\alpha$ much lower than in the surrounding AGN-ionized gas; in Hanny's
Voorwerp, both signatures occur in the same places. We require  cluster candidates to be detected in both our continuum bands, to reject remaining cosmic-ray 
artifacts and reduce confusion with fine structure in dust lanes. Single clusters will be essentially pure PSF images at these redshifts, although blending of
adjacent clusters (as well as incomplete correction of CTE effects in the images) leads us to consider slightly resolved objects as well (with measurements
described in \S \ref{s-starclusters}).
None of the target galaxies has a rich cluster population. Our color information is somewhat limited
in precision by the modest wavelength baseline between our continuum filters, which were chosen primarily for the best match to emission-line bands. Except
in the inner regions of the galaxies where Poisson noise increases, the $5 \sigma$ detection threshold is about $m_{621}=25.5$, $m_{763}= 24.7$
for measurements within a 0\farcs 4 radius. For faint clusters or those with close neighbors, we use $r=0.2${\arcsec} and a point-source correction
to match the zero points given for 0.4\arcsec (http://www.stsci.edu/hst/wfc3/phot\_zp\_lbn). To reduce contamination from foreground stars or background
galaxies, we accept only cluster candidates with $(R-I) < 0.9$, and computed absolute magnitude $M_{621}$ fainter than -15. No such
objects are found in NGC 5252 and SDSS 1510+07. Photometric values for candidate clusters are given in Table \ref{tbl-clusterphot}.

\begin{figure*} 
\includegraphics[width=140.mm,angle=00]{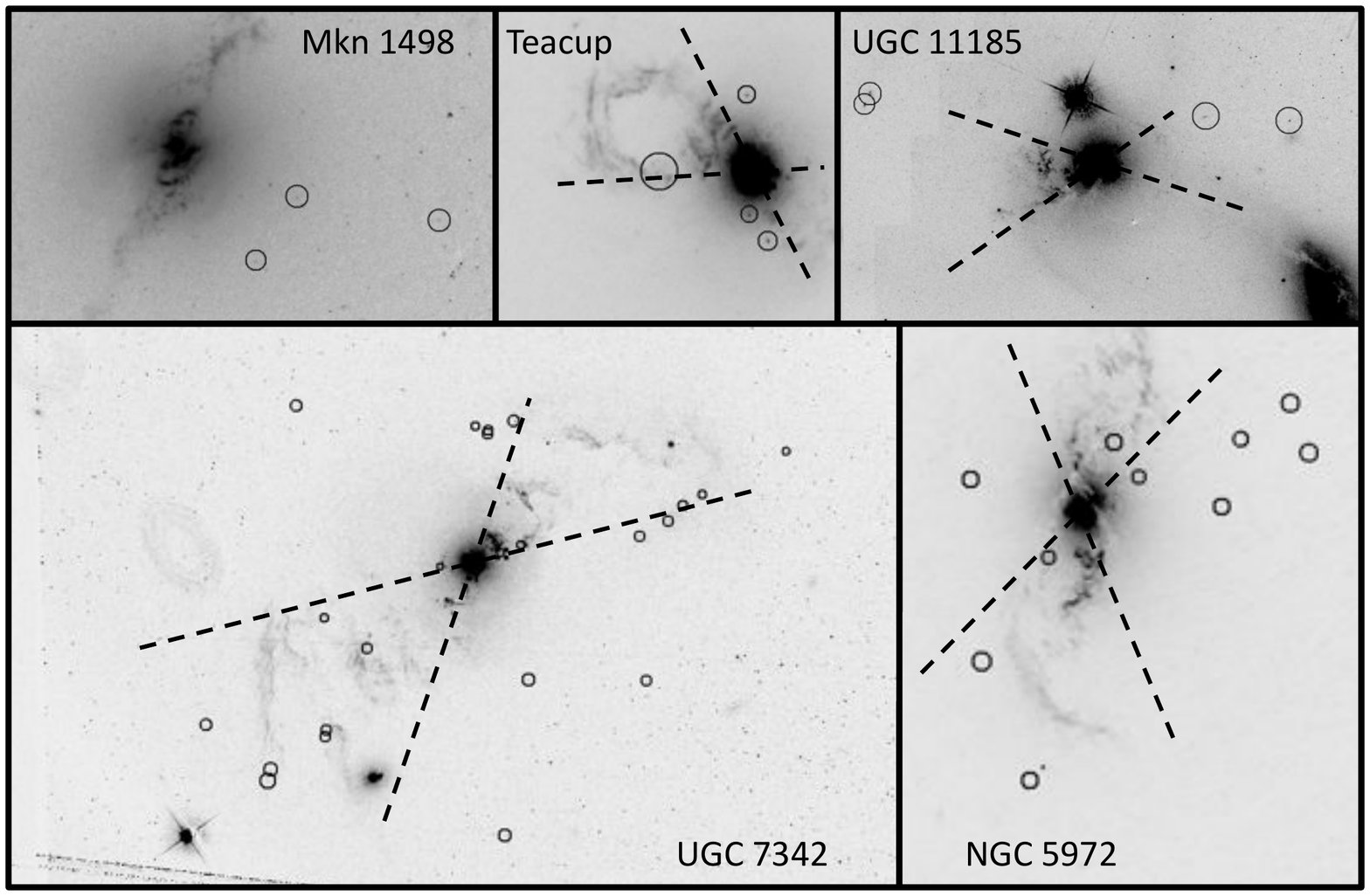} 
\caption{Spatial distributions of star-cluster candidates. North is at the top in each panel; to show both stellar and gaseous structures, the base images here are the continuum 
plus \oiii mixes used in Fig. \ref{fig-overview1}. Dashed lines mark approximate boundaries of ionization cones as inferred from the \oiii/H$\alpha$ ratio.
The large circle in the Teacup system marks a likely background galaxy, which might otherwise be interpreted as $\approx 6$ blue clusters surrounding a redder
object.} 
\label{fig-clusterdist}
\end{figure*}

None of these cluster candidates shows an \oiii/H$\alpha$ signature as we saw in Hanny's Voorwerp. Most of them are found too far from
strong line emission (and those which are close to emission filaments do not show the ratio reversal that indicates a change to ionization by the UV continuum of
hot stars). We do see a broad association between regions of high \oiii/H$\alpha$ (``ionization cones") and the cluster candidates, which we might
speculate indicates a long-term connection between the AGN and cluster formation, as shown in Fig.  \ref{fig-clusterdist}.
The cone angles in projection range from 57-68$^\circ$, so a random distribution would have about 1/3
of the points within the cones. Among all cluster candidates, we see a small excess over this (20/43), although the significance is marginal and could be changed by 
small variations in the boundaries in UGC 7342 and NGC 5972. Any outflow-induced star formation within the emission regions in these galaxies
must be a minor effect, or have been operating over long times compared to the light-travel times sampled by the observed gas.

{\bf Mkn 1498} has 3 candidate clusters, all to the SW of the nucleus and roughly perpendicular to the axis of the extended ionized gas. All are far from
the ionized gas.

{\bf NGC 5972} shows 9 cluster candidates within projected radius 30{\arcsec} (17 kpc), with no obvious preferred alignment. Their absolute magnitudes range
from -10.2 to -11.3, with a color range centered near $(R-I) = 0.30$. Four of these nine lie projected within the ionization cones

Cluster candidates exist in {\bf SDSS 2201+11}, but are so close to the dust lanes that precise photometry is greatly limited by fluctuations in background and attenuation,
so they are not shown in the montage or listed in Table \ref{tbl-clusterphot}.

In the {\bf Teacup} host galaxy, we find several candidate clusters within the inner 5\arcsec, mostly concentrated in a clump about 3\farcs 5 E  of the nucleus or 
roughly aligned with the dust lane. We measure their fluxes
after subtracting a smooth model for the galaxy, and use local statistics to evaluate photometric errors. The prominent clump may in fact be a background
galaxy, since its structure of a redder core surrounded by blue knots is reminiscent of many disk galaxies seen in the emitted UV; we measure
total values with $r=1${\arcsec} of $m_{763}=21.05 \pm 0.03$, $m_{632}=21.61 \pm 0.03$, $(R-I)_0 = 0.61 \pm 0.05$.   All these candidates occur
in regions of lower-ionization gas outside the ionization-cone boundaries. 
The two bright objects outside the dust lane have  $M_R \approx -14$,  $(R-I)_0 \approx 0.30$.

In {\bf UGC 7342}, we find 23 candidate clusters in our color and luminosity range, from $M_R = -13$ and fainter. Five of these
are in the prominent tidal tail to the NW, and three are projected in the inner 6{\arcsec} (5.5 kpc). Sixteen are quite blue, near
$R-I = 0.30$. As with UGC 11185, the 5 cluster candidates in the tidal tail lie 
near the projected boundary of the large-scale ionization cone. The clusters here are more luminous than in the
other systems, with three brighter than $M_R = -12$ and all the detections brighter than $M_R = -11$. There is
marginal evidence for a bimodal cluster population with peaks near $(R-I)_0 = 0.3$ and 0.75. Half the
candidates (12/23) are within the ionization cones, albeit in may cases also within stellar tidal features.

{\bf UGC 11185} is seen at a low galactic latitude, so the potential contamination by bright foreground stars makes the bright magnitude cut for
candidates more important.  This leaves four distinct cluster candidates within 30{\arcsec} (24 kpc) of the nucleus, one of which has at last three subcomponents at the
0\farcs 15 level. This object and an close neighbor lie on the inner edge of the tidal loop to the E of the nucleus; the
other two lie almost opposite this complex on the W side of the system. All four objects are projected just within the
angular span of the ionization cone seen closer to the AGN - two in a stellar tidal stream, and two possibly associated with the companion
spiral galaxy to the WSW. These are potential cases of outflow-induced star formation.

We evaluate the surface density of foreground and background objects which could potentially masquerade as star clusters, using the
outer parts of the WFC3 fields (except for NGC 5972, this used the CCD not including the galaxy center). Performing similar photometry on
objects which are unresolved or marginally resolved, excluding obvious background galaxies, we derived object counts in bins of $R_{621}$ and
$(R-I)$ for $R_{621} < 25.0$ and requiring detections on both filters (as for the cluster candidates). Based on the area covered by our target galaxies, 
the expected number of unrelated objects within our color range is as large as 1 only for NGC 5972 and UGC 7342, in the range $R_{621}=24-25$. Contamination
does not significantly affect our statistics of cluster candidates. Most potentially contaminating objects are quite red; about 75\% are in the range $(R-I)= 0.7 - 2.5$.

None of these galaxies is rich in young clusters, despite the strong evidence of interactions and mergers, and
substantial supplies of gas and dust. We might speculate that
AGN episodes could have previously quenched star formation, or that there is some
strong selection in favor of bulge-dominated systems to have the giant ionized clouds. In the former case, dust lanes might not only be
plausible locations for molecular clouds, but could play a role in shielding such clouds from radiative interference by the AGN.

Our color cut is generous for ages and reddenings of candidate clusters; in comparison, the Starburst 99 models \citep{Starburst99} show simple stellar populations
briefly becoming as red as $(R-I)=0.9$ at ages near $8 \times 10^6$ years for high upper mass limits, as red supergiants are most important in the integrated
light (so this is the time of peak near-IR luminosity), but otherwise remaining bluer than $(R-I)=0.6$. With many candidates far from the galaxy cores, we would expect only local reddening except near the dust lanes.

\begin{deluxetable}{lrcccc}
\tablecaption{Candidate star clusters \label{tbl-clusterphot}}
\tablewidth{0pt}
\tablehead{
\colhead{Galaxy} & \colhead{ID} &   \colhead{$\alpha, \delta$ (2000)} & 	\colhead{$m_{763}$} & \colhead{$m_{632}$} & \colhead{$(R-I)_0$}  }
\startdata
Mkn 1498 & 1 & 16:28:03.675 +51:46:26.43 & $ 25.26 \pm 0.16$ &  $25.33 \pm 0.10$ & $ 0.08 \pm 0.18$ \\
	 & 2 &16:28:03.484 +51:46:29.17 & $25.10 \pm 0.15$ &  $25.03 \pm 0.08$ & $ -0.07 \pm 0.17$ \\
	& 3 &16:28:02.826 +51:46:28.25 & $ 24.75 \pm 0.10$ & $ 25.33 \pm 0.09$ & $ 0.63 \pm 0.15$ \\
NGC 5972 & 1 & 15:38:54.374 +17:01:29.88  & $24.46 \pm 0.11$ & $ 24.61 \pm 0.08$ & $ 0.16 \pm 0.14$ \\
	& 2 & 15:38:53.830 +17:01:37.16 & $24.49 \pm 0.11$ & $ 25.05 \pm 0.12$ & $ 0.67 \pm 0.16$ \\
	& 3 & 15:38:53.979 +17:01:40.28 & $24.81 \pm 0.15$ & $ 25.10 \pm 0.12$ & $  0.32 \pm 0.19$ \\
	& 4  & 15:38:53.188,+17:01:40.49 & $24.46 \pm 0.07$ &  $ 24.74 \pm 0.07 $ & $0.30  \pm 0.11$ \\
	& 5  & 15:38:53.315 +17:01:34.56 & $24.87 \pm 0.11$ &  $ 25.00  \pm 0.07 $ & $0.14  \pm 0.13$ \\
	& 6  & 15:38:52.760 +17:01:39.35 & $24.22 \pm 0.06$ &  $ 24.64  \pm 0.05 $ & $0.45  \pm 0.10$ \\
	& 7  & 15:38:52.876 +17:01:43.77 & $24.94 \pm 0.12$ &  $ 25.33  \pm 0.10 $ & $0.42  \pm 0.16$ \\
	& 8  & 15:38:54.893 +17:01:37.06 & $23.43 \pm 0.03$ &  $ 24.58  \pm 0.05 $ & $0.17  \pm 0.07$ \\
	& 9  & 15:38:54.819 +17:01:20.50 & $23.85 \pm 0.04$ &  $ 24.26  \pm 0.04 $ & $0.45  \pm 0.07$ \\
	& 10  & 15:38:54.518 +17:01:09.83 & $25.69 \pm 0.23$ &  $ 26.42  \pm 0.26 $ & $0.79 \pm 0.35$ \\
Teacup & 1 &14:30:29.910 +13:39:14.43 &  $23.47 \pm 0.04$ & $ 23.81 \pm 0.03$ &  $0.37 \pm 0.05$ \\
	& 2    & 14:30:29.857 +13:39:09.02 & $23.57 \pm 0.07 $ & $23.84   \pm 0.05 $ & $  0.30 \pm 0.09$ \\
	& 3     & 14:30:29.907 +13:39:09.95 & $24.06 \pm 0.05$ & $ 24.44  \pm 0.08 $ & $  0.42 \pm 0.12$ \\
	& 4     &  14:30:30.137 +13:39:11.51 & $22.93 \pm 0.02 $ & $23.45   \pm 0.02  $ & $ 0.57 \pm 0.03$ \\
	& 5   & 14:30:30.131 +13:39:11.31 & $23.47  \pm 0.03 $ & $23.66  \pm 0.03 $ & $  0.21 \pm 0.05$ \\
	& 6   & 14:30:30.155 +13:39:11.44 & $24.27 \pm 0.06 $ & $24.54  \pm 0.06  $ & $  0.30 \pm 0.09$ \\
	& 7     & 14:30:30.151 +13:39:11.90  & $23.48\pm 0.04 $ & $ 23.86  \pm 0.03  $ & $  0.42 \pm 0.05$ \\
UGC 7342 & 1 & 12:18:16.507 +29:15:25.87 & $25.12  \pm 0.15  $ & $25.35  \pm 0.09 $ & $0.25  \pm 0.17$ \\
	& 2 & 12:18:17.298 +29:15:20.77 & $24.05  \pm 0.06 $ & $24.16  \pm 0.04 $ & $0.12  \pm 0.07$ \\
	& 3 & 12:18:17.517 +29:15:19.31 & $24.87  \pm 0.12 $ & $25.08  \pm 0.08 $ & $0.21  \pm 0.14$ \\
	& 4  & 12:18:17.592 +29:15:17.80 & $24.55  \pm 0.09 $ & $25.12  \pm 0.08 $ & $0.63  \pm 0.12$ \\
	& 5   & 12:18:17.859 +29:15:15.93 & $24.01  \pm 0.05 $ & $24.45  \pm 0.05 $ & $0.48  \pm 0.07$ \\
	& 6   & 12:18:19.215 +29:15:27.95 &  $24.60  \pm 0.08 $ & $25.14  \pm 0.07 $ & $0.59  \pm 0.13$ \\
	& 7  & 12:18:19.224 +29:15:28.41 & $24.04  \pm 0.05 $ & $24.82  \pm 0.06 $ & $0.78  \pm 0.09$ \\
	& 8   & 12:18:19.335 +29:15:28.77 &  $25.12  \pm 0.13 $ & $25.35  \pm 0.09 $ & $0.25  \pm 0.16$ \\
	& 9    & 12:18:20.935 +29:15:31.25 &  $24.50  \pm 0.08 $ & $24.72  \pm 0.05 $ & $0.24  \pm 0.11$ \\
	& 10    & 12:18:20.688 +29:15:06.16 &  $25.06  \pm 0.13 $ & $25.26  \pm 0.08 $ & $0.22  \pm 0.15$ \\
	& 11   & 12:18:20.298 +29:15:02.67  & $24.89  \pm 0.11 $ & $25.46  \pm 0.10 $ & $0.63  \pm 0.15$ \\
	& 12    & 12:18:21.752 +29:14:53.63 & $24.65  \pm 0.09 $ & $24.71  \pm 0.05$ & $ 0.07  \pm 0.12$ \\
	& 13    &  12:18:20.671 +29:14:52.95 &$24.70  \pm 0.09 $ & $24.87  \pm 0.06 $ & $0.19  \pm 0.13$ \\
	& 14    & 12:18:20.687 +29:14:52.43 &  $24.91  \pm 0.11 $ & $25.21  \pm 0.08 $ & $0.33  \pm 0.14$ \\
	& 15    & 12:18:21.168 +29:14:48.42 &  $24.65  \pm 0.09 $ & $24.73  \pm 0.05 $ & $0.09  \pm 0.12$ \\
	& 16    & 12:18:21.203 +29:14:47.07  & $24.46  \pm 0.07 $ & $24.74  \pm 0.05 $ & $0.31  \pm 0.10$ \\
	& 17    & 12:18:19.065 +29:14:40.70 &  $24.43  \pm 0.17 $ & $25.12  \pm 0.07 $ & $0.20  \pm 0.18$ \\
	& 18    & 12:18:18.851 +29:14:59.02 & $24.44  \pm 0.07 $ & $24.74  \pm 0.05 $ & $0.33  \pm 0.10$ \\
	& 19    & 12:18:17.793 +29:14:58.70 & $24.08  \pm 0.05 $ & $24.68  \pm 0.05 $ & $0.66  \pm 0.07$ \\
	& 20    & 12:18:19.048 +29:15:29.21  &  $23.56  \pm 0.03 $ & $23.84  \pm 0.01 $ & $0.30  \pm 0.03$ \\
	& 21   & 12:18:19.650 +29:15:12.25  &  $24.39  \pm 0.16 $ & $25.03  \pm 0.11 $ & $0.70  \pm 0.19$ \\
	& 22   & 12:18:19.101 +29:15:13.86  & $24.56  \pm 0.16 $ & $24.70  \pm 0.10 $ & $0.16  \pm 0.18$ \\
	& 23   &  12:18:18.926 +29:15:14.85 & $25.15  \pm 0.18 $ & $25.19  \pm 0.10 $ & $0.04  \pm 0.20$ \\
UGC 11185 & 1 & 18:16:13.763 +42:39:44.60 & $ 22.06  \pm 0.02 $ & $22.64  \pm 0.02 $ & $0.53  \pm 0.03$ \\
	& 2  & 18:16:13.811 +42:39:43.45& $22.58  \pm 0.03 $ & $22.84  \pm 0.03$ & $ 0.29  \pm 0.03$ \\
	& 3  &18:16:09.773 +42:39:41.80  & $22.37  \pm 0.03 $ & $23.00  \pm 0.02 $ & $0.69  \pm 0.04$ \\
	& 4   & 18:16:10.575 +42:39:42.21& $22.87  \pm 0.04$ & $ 23.44   \pm 0.04 $ & $0.68  \pm 0.06$ \\
\enddata
\end{deluxetable}

\section{Gas kinematics\label{sec-gaskin}}

For the large-scale emission structures, the main feature of the velocity fields  in these systems is that most appear
rotationally dominated, following
ordinary rotation curve patterns with only modest local displacements or velocity dispersions. This stands in strong contrast to the
EELRs around radio-loud quasars (\citealt{SFC2006}, \citealt{FuStockton2009}), which show emission features over
ranges 1000-2000 km s$^{-1}$ (possibly resulting from interactions with the jets).  For the objects with Fabry-Perot velocity fields, we can 
extend this conclusion to the full systems rather than sampling along single slits.

To characterize local departures from ordered motion, we consider residuals from median-smoothed velocity fields in the BTA
data. For each system we construct a smoothed velocity field by taking the median of values within $9 \times 10${\arcsec}
($13 \times 25$ pixels) boxes, with the short axis along the overall velocity gradient. On comparison of the velocity
and velocity-dispersion maps (Fig. \ref{fig-btamaps}), we find that almost all
values $\sigma_v > 100$ km s$^{-1}$ occur in regions of low surface brightness or strong gradients, where both beam smearing
and seeing-match issues are strongest. 

{\bf NGC 5252} remains an outlier, in that its velocity gradient is almost perpendicular to the ionization cone (as seen in the H I data
of  \citealt{PrietoFreudling} and the optical spectra of \citealt{Acosta1996} and \citealt{morse}).
The velocity field in this case is not purely planar rotation, as seen from the inflections in its contours. The only locally high velocity
dispersion is at the AGN itself.

The velocity field appears continuous in {\bf  NGC 5972} into the detached emission regions outside the
ACS narrowband images. The line profiles are asymmetric, possibly reflecting projected of material at very different
line-of-sight locations, within 3{\arcsec} north and south of the nucleus, in the sense that additional components are present closer to
systemic velocity than the main rotation pattern. This is not seen along the minor axis, so this velocity structure is not due to seeing effects.
A similar effect occurs in the longslit data along PA 165$^\circ$: the velocity shows a displacement toward the systemic value
11{\arcsec} N and S of the nucleus, where the dust attenuation is strongest. This spectrum shows stellar absorption in the Mg+MgH 
region in the inner 6{\arcsec} sufficient to show that the stellar rotation is in the same sense as we see in the gas.

In {\bf SDSS 2201+11}, the southern emission-line cloud may be two projected in overlap differing in radial velocity by
$\sim 50$ km s$^{-1}$. There are distinct velocity
components separated by about 7{\arcsec} east-west, with a local increase in $\sigma_v$ at their projected interface. 
The higher-velocity component corresponds to distinctly more diffuse material on the western side of this
structure. The velocity contours in this system are not purely tangential; the cloud edges away from the main dust disk plane
have smaller rotational velocity, which may be explained by a warped disk model. The long-slit rotation slice shows
a nearly undisturbed, typical rotation curve with circular velocity near 235 km s$^{-1}$. This spectrum shows stellar absorption in the Mg+MgH 
region in the inner 20{\arcsec} sufficient to show that the stellar rotation is in the same sense as we see in the gas.

The emission-line mapping of the {\bf Teacup} system reported by \cite{Harrison} shows a localized region of large blueshifts
within 1\arcsec SW of the AGN, with projected velocity offset 760 km s$^{-1}$. Their radio mapping shows continuum
emission coincident with part of the large loop of emission to the west, suggesting that this feature also marks an outflow
(perhaps entrained material around the result of a previous radio outburst).

In {\bf UGC 7342}, there is an outlying region 100 km s$^{-1}$ redward of the smooth velocity field seen 8.5{\arcsec} northwest of the nucleus.
The gas distribution in this system is perpendicular to the circumnuclear dust lanes, and so is the velocity gradient; this might
be an extreme example of a warped and precessing disk, with damping significant in bringing the material toward the plane of the
main galaxy potential.
Important orientation angles are compared in Table \ref{tbl-angles}. The kinematic major axis, perpendicular to the kinematic line of nodes (LON), is often 
(but not always) near the major
axis of the emission clouds.

In {\bf UGC 11185}, there are three \oiii velocity components on the east side of the system. The higher-velocity extensive system is associated in the Fabry-Perot 
data with the conical or triangular region bright in the HST \oiii image. This component dominates the velocity fits where it is brightest, and blending in this
region may account for some of the velocity structure seen by \cite{mnras2012}; on both sides the
lower--velocity ``systemic" curve shows a smoother, albeit still disturbed and asymmetric, form. The higher-velocity emission could trace an outflow, or
ionization of a complex tidal tail projected against the inner part of the galaxy. The high abundances in the UGC 11185 extended clouds may argue 
for an outflow interpretation.

\begin{deluxetable}{lcccc}
\tablecaption{Structural and kinematic axes \label{tbl-angles}}
\tablewidth{0pt}
\tablehead{
\colhead{Galaxy} &	\colhead{Gas major axis [$^\circ$]} & \colhead{Kinematic LON [$^\circ$]} & \colhead{Kinematic major axis [$^\circ$]} & \colhead{Dust lanes [$^\circ$]}}
\startdata
NGC 5252	&	161			&	6		&	104\\
NGC 5972	&	169			&	96		&	6 & 169 \\
UGC 7342	&	131			&	38		&	128 &  29\\
SDSS 2201+11 &	21			&	118		&	28   & 173  \\ 
\enddata
\end{deluxetable}

\begin{figure} 
\includegraphics[width=70.mm,angle=0]{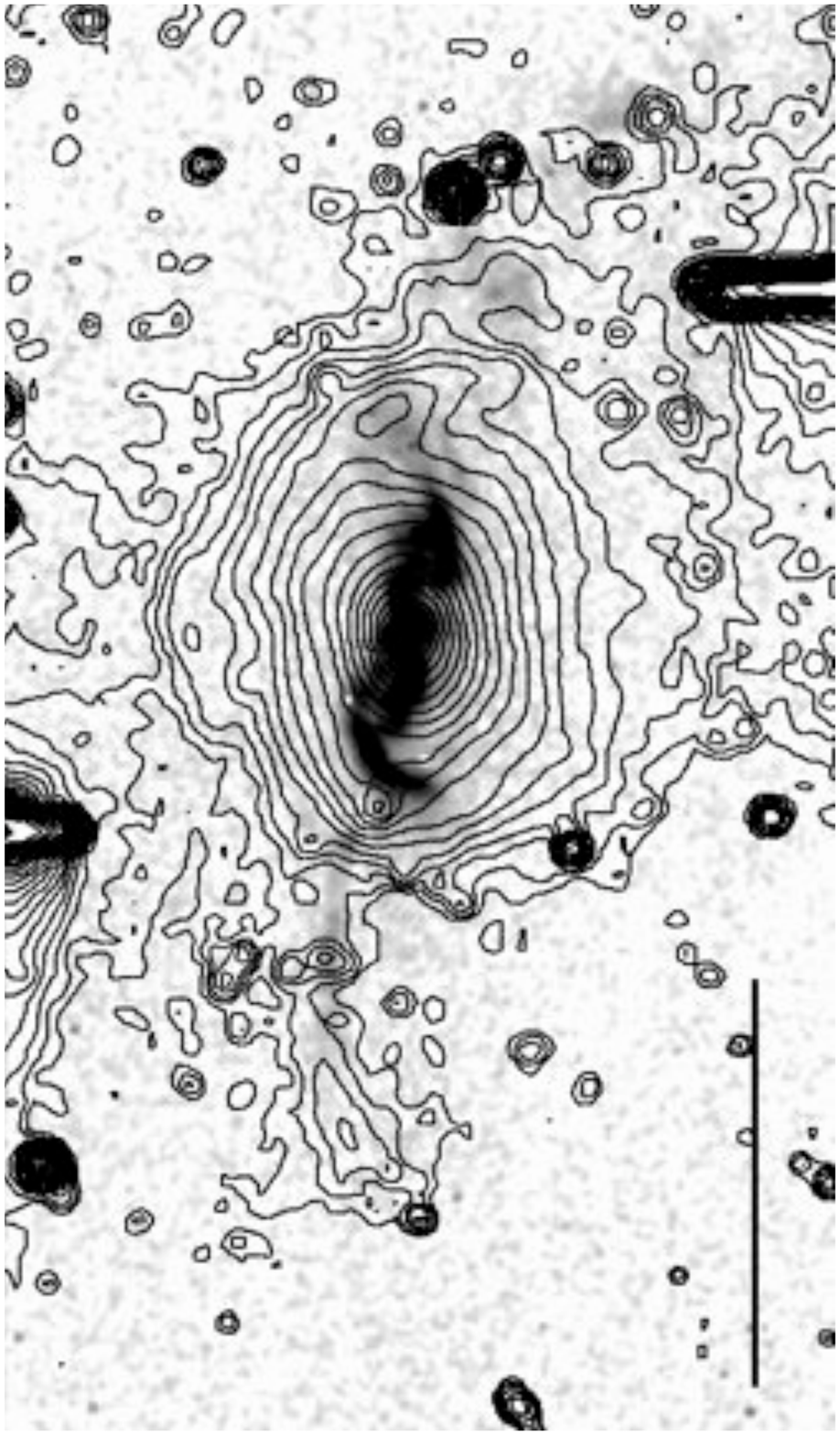} 
\caption{Stellar and gas tails in NGC 5972, from data with the KPNO 2.1m telescope. Contours show logarithmic intensity of 
a $V$ continuum image, corrected for
the \oiii line emission, and the grey scale shows continuum-subtracted \oiii. Especially in the southern (lower) tail, the line emission broadly
follows the stellar tail, but has an offset center line. Residuals from bright stars appear at opposite corners of the view. The scale bar
shows 1 \arcmin.} 
\label{fig-ngc5972kpno}
\end{figure}

The ionized-gas distributions in the tidal tails broadly match the stellar tails seen in continuum images, but there are small-scale
offsets that might in principle either show genuine offsets in gas and star distributions (which naturally results from the different radial
and velocity distributions of these components in the original disk before tidal disturbance) or show us three-dimensional structure
in the tidal debris weighted by the distribution of ionizing radiation for the gas. These differences are most clearly seen in the outer parts
of NGC 5972 (mostly beyond the field of the HST emission-line images), as shown in Fig. \ref{fig-ngc5972kpno}.

\section{Gas fine structure}


As was done for IC 2497 and Hanny's Voorwerp by \cite{HSTHV}, we compare the pixel-by-pixel surface brightness in H$\alpha$ to the ionization
parameter $U$ to test for the structure in the ionized clouds beyond the resolution limit of HST data. We can map the observed \oiii/H$\alpha$
ratio to changes in $U$ because these clouds are photoionized by a distant source, so that the continuum shape is fixed and we can isolate
regions with a small range in (projected) radius from the AGN. Under these conditions, if a resolution element samples a smooth distribution of
matter, the surface brightness in a recombination line will scale as $SB \propto {n_e}^2$, and $U$ will scale as ${n_e}^{-1} \propto SB^{-1/2}$.
At these ionization and density levels, \oiii/H$\alpha$ scales in a nearly linear way with $U$ (e.g. \citealt{FerlandNetzer}). Our test thus uses the 
relation between H$\alpha$ SB and
\oiii/H$\alpha$ for individual features in the ionized clouds ; significant departures from the $SB^{-1/2}$ scaling indicate that the conditions above are not 
satisfied. In Hanny's Voorwerp, the departures were in the sense that brighter pixels had $U$ decrease more slowly than predicted; a straightforward interpretation is
that there is unresolved structure, so that regions of higher surface brightness have a larger number of unresolved clumps and filaments rather than
higher mean density.

\begin{figure*} 
\includegraphics[width=148.mm,angle=270]{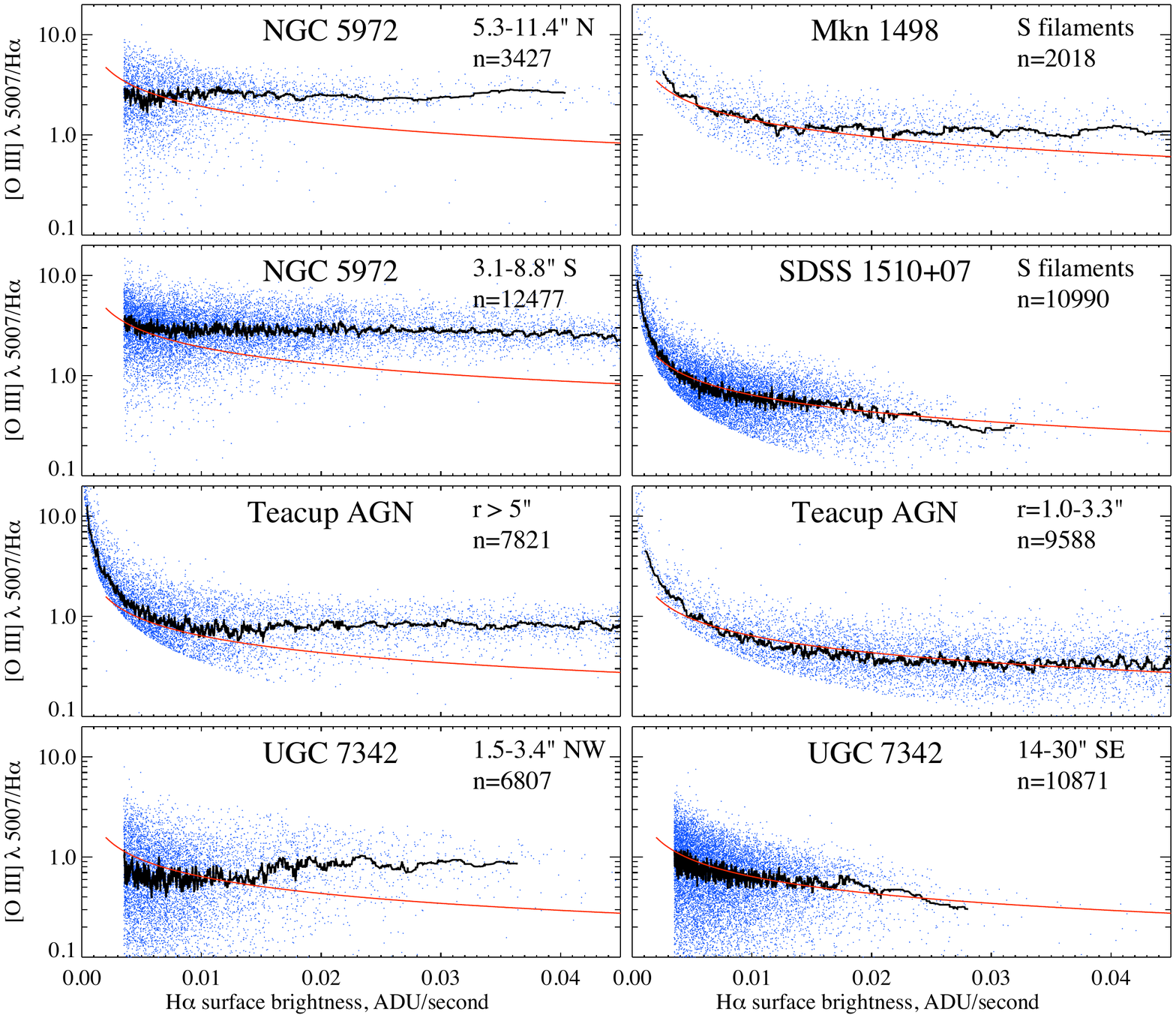} 
\caption{Examples of poor and good fits of the H$\alpha$ surface brightness -- ionization-parameter relation to the expectation if H$\alpha$ 
surface brightness represents constant $n_e$ within each resolution element, shown by the schematic curve for each panel. Left column, poor fits indicating
significant unresolved density structure. Regions in the right column have relations that are well fit by the simple expectations. Points shown are
masked at $2  \sigma$ error is H$\alpha$, and a $3 \times 3$-pixel median filter was used to reject residual bad pixels. Each panel is labelled with the
number of pixels shown. For each data set, a 51-pixel running median is shown to illustrate the behavior of the distributions.} 
\label{fig-hasbplots}
\end{figure*}

In these systems, this test gives mixed results. As shown in the samples in Fig. \ref{fig-hasbplots}, some regions (each including pixels at nearly the 
same projected distance
from the nuclei) are reasonable fits to the expected behavior if the HST images resolve most of the structure in the ionized gas. Others show a much
weaker dependence of ionization parameter on H$\alpha$ surface brightness than expected, suggesting that additional fine-scale structure exists beyond the HST
resolution limit. The latter cases include the inner filaments in UGC 7342, radial samples in the loop of the Teacup system, the outer parts of the radial
clouds in Mk 1498, and the inner bright filaments in NGC 5972.
In a very broad way, clouds at the largest radii, often with lower mean surface brightness, are more likely to follow the simple $n_e - U$ behavior for
a smooth distribution of material. Small features cannot be evaluated with this test if there are too few pixels to establish the relationship.

In several of these galaxies, the ionized gas shows coherent filaments extending several kpc, sometimes in parallel sets. Good examples
are seen in SDSS 1510+07, extending as much as 7.8{\arcsec} (7.5 kpc). These are barely resolved, with widths around 0.15{\arcsec} (140 pc). This combination of
length and narrowness may indicate confinement; even at the thermal velocity for H (17 km s$^{-1}$ rms) associated with a typical electron temperature 12,000 K, 
such narrow filaments would disperse much faster than the dynamical timescales $\approx 10^8$ years given by the rotation velocity and size of
the filaments. One could picture the filaments constantly being regenerated by instabilities or being preserved by confinement, both of which may
suggest magnetic fields which are dynamically important on local scales for the diffuse gas and ordered over kpc scales. Such fields could shape the 
geometries of the clouds on large scales, both in filaments and the twisted ``corkscrew" forms which are suggested in several of these galaxies
(such as UGC 7342). 

Some of these systems have extended radio emission (notably NGC 5972, Mkn 1498, and on smaller scales the Teacup). In these, our 
conjecture about magnetic confinement might become testable through sufficiently sensitive and high-resolution Faraday rotation measurements.
The nature of the emission clouds -- externally photoionized sections of tidal debris - introduces the possibility that such fine structure (and possibly
embedded magnetic fields) could be a common feature of gas in tidal features, currently observable only when we see optical emission lines
from the ionized portions. 

\section{Origin of the ionized gas}

Since the ionized clouds we examine are (by selection) more than 10 kpc from the AGN, they are unlikely to be regions of ordinary disk interstellar gas.
Such distant material might represent outflows driven by the AGN, or recently-acquired gas in tidal tails or streams. The morphology, kinematics,
and abundances in the gas all bear on this question.

Tidal distortions are present in the stellar morphologies of 7/8 galaxies in this sample; in the remaining one, NGC 5252,  we see dust out of the galaxy plane, and
H I data show a rough polar ring of
gas with no detected stellar counterpart. Where the large-scale kinematic pattern is dominated by ordered rotation and the abundances are sigificantly
lower than at the nucleus, the case is strong that we are seeing photoionized tidal debris (although it is possible that outflows may be important
in a small region around the AGN itself). This applies to UGC 7342, with some of the lowest abundance values we observe all across the
ionized clouds;  to SDSS2201+11; and to Hanny's Voorwerp itself (in which H I data from \citealt{Josza} show the ionized gas to be part of a much larger, 
kinematically coherent structure). NGC 5252 falls in this category as well; the outer emission regions have significantly lower abundances than seen in the
inner few arcseconds. Since we trace kinematics using emission lines, we do not always know whether the sense of rotational order matches that
of the stars; in NGC 5972, we detect absorption lines well across enough of the galaxy to know that the senses do match.

The more complex kinematics of the Teacup system and UGC 11185, with greater velocity ranges and localized multiple components, make a case
for outflows playing a more important role. In these systems, the extended-gas abundances are close to those in the respective nuclei, likewise
consistent with an outflow. Kinematically, it s also possible that the second emission system seen in the eastern part of UGC 11185 is tidal
debris projected against the host galaxy, but then one might expect lower abundances.

An additional signature of a small-scale outflow would be a velocity mismatch between the AGN emission and the interpolated central value of the
velocity field (or slice along the slit location). We can examine this point for some of our sample.
NGC 5252 shows a rotation of velocity contours by $\approx 120^\circ$ in the inner 10\arcsec, with a separate region in the inner 2\arcsec at about $90^\circ$ to the outer velocity field. Each of these might be as well described as a counterrotating region or a
radial perturbation  of order 70 km s$^{-1}$ to a global pattern.  In NGC 5972 and SDSS 2201, any specifically nuclear velocity offset must be at a level $< 30$ km s$^{-1}$.

\section{Summary}

We have used HST images and a range of supporting data to study the host galaxies of extended emission-line regions (EELRs) whose AGN are
insufficient to photoionize them to the observed levels, implying significant fading of the AGN radiation on timescales $< 10^5$ years. The kinematics of the
ionized gas are dominated by rotation. Outflows are not globally
important in these objects, unlike the EELRs in some radio-loud AGN hosts. We see several
clear cases of warped and precessing disks of gas and dust, and the gas is broadly associated with stellar tidal tails and irregular dust 
distributions in others. Each of the 8 objects in our sample shows evidence of a merger or strong interaction; these EELRs are largely photoionized
tidal debris. With only the single possible exception of Mkn 78, the sample of comparable nearby EELRs where the AGN are energetically
capable of lighting them up are likewise in interacting systems; this makes sense, given that the light-travel timescales we
can study are much shorter than the dynamical times for evolution of tidal distortions (broadly, $10^5$ versus $10^9$ years). 
Both the ionizing radiation from the AGN and large reservoirs of distant neutral gas are required to see EELRs such as these, giving a strong selection
effect in favor of AGN with tidal debris; episodic
behavior of the AGN similar to that we infer for these objects could be equally common in noninteracting galaxies but we would seldom be in a position to observe it. One
avenue might be use of neighboring galaxies' IGM to measure or limit the AGN radiation reaching them; we have conducted a pilot survey for such cases
(Keel et al,. in preparation).

The kinematics  in the extended clouds show some instances (SDSS 2201+11, UGC 7342, Hanny's Voorwerp itself)
of material strongly dominated by
ordered rotation. 
These generally show low (nitrogen/oxygen) abundances in the gas, consistent with an external tidal origin in less
massive and lower-abundance systems. In some systems (the Teacup AGN, UGC 11185) , we see multiple superimposed velocity systems and localized high velocities
(although lower than previously found in many radio-loud AGN), 
sometimes accompanied by near-solar abundances across the clouds; these are candidates for outflows at least affecting the clouds.

In detail, the gas distributions show a variety of forms - giant loops, long filaments, and partial shells. The long filaments, and
in some structures a breakdown of the expected relation between H$\alpha$ surface brightness and ionization parameter in the gas,
may indicate a role for magnetic fields, possibly pre-existing in the tidal tails rather than linked to the AGN. Since detection these structures
requires the angular resolution of HST (and in some cases lies beyond this limit), other tidal tails may be similarly structured although we cannot
yet observe at these levels with other techniques (such as H I mapping). 

We find only marginal evidence for association between the locations of emerging UV flux (as traced by the ionized gas) and
young star clusters; any role for triggering star formation by outflows in these cases must be very limited. None of the galaxies
has extensive star formation; one is an S0, one an SB0, and the rest strongly bulge-dominated even when tidal tails are present.
			
\acknowledgments

This work was supported by NASA through STScI grants HST-GO-12525.01-A and -B.
Some of the data presented in this paper were obtained from the Mikulski Archive for Space Telescopes (MAST). 
This research has made use of NASA's Astrophysics Data System, and
the NASA/IPAC Extragalactic Database (NED) which is operated by the Jet Propulsion Laboratory, California Institute of Technology, under contract with the National Aeronautics and Space Administration. We thank Linda Dressel for advice on setting up the observations, especially
reducing intrusive reflections. Giles Novak provided useful pointers on magnetic fields in the ISM. Identification of this galaxy sample
was possible through the efforts of nearly 200 Galaxy Zoo volunteers; we are grateful for their contributions, and thank once more the
list of participants in \cite{mnras2012}. This work was partly supported  by the ÔActive Processes in Galactic and Extragalactic ObjectsÕ basic research programme of the Department of Physical Sciences of the RAS OFN-17.  The observations obtained with the 6-m telescope of the SAO of the RAS were carried out with the financial support of the Ministry of Education and Science of the Russian Federation (contract no. N14.619.21.0004,  the project  RFMEFI61914X0004).

W.P. Maksym is grateful for support by the University of Alabama Research Stimulation Program.
V.N. Bennert acknowledges assistance from a National Science Foundation (NSF) Research at Undergraduate Institutions (RUI) grant AST-1312296. Note that findings and conclusions do not necessarily represent views of the NSF.
C. J. Lintott
acknowledges funding from The Leverhulme Trust and the STFC Science in Society Program. 
A. Moiseev is also grateful for the financial support of the ÔDynasty' Foundation and a grant from the President of the Russian Federation (MD3623.2015.2). 
K. Schawinski was supported  by a NASA Einstein Fellowship at Yale, and gratefully acknowledges support
from Swiss National Science Foundation Grant PP00P2 138979/1. Galaxy Zoo was made
possible by funding from a Jim Gray Research Fund from Microsoft and The Leverhulme Trust.

This publication makes use of data products from the Wide-field Infrared Survey Explorer, which is a joint project of the University of California, Los Angeles, and the Jet Propulsion Laboratory/California Institute of Technology, funded by the National Aeronautics and Space Administration.

{\it Facilities:} 
\facility{HST (ACS, WFC3, WFPC2)},
\facility{BTA}, \facility{Shane (Kast Double spectrograph)}, \facility{SARA},
\facility{WISE}, \facility{KPNO:2.1m}



\appendix

\section{Appendix: Object properties}

For more convenient reference, we collect here descriptions of our results organized by  galaxy, rather than by technique as detailed earlier.

{\bf Mkn 1498} has the most prominent broad-line region seen in our sample, perhaps a clue to the orientation of its
AGN structures. There are multiple, partial circumnuclear rings of ionized gas at radii 0.5--1.6{\arcsec} (0.6-1.8 kpc).
While the starlight distribution of Mkn 1498 is nearly symmetric, it does include low-contrast fans or spirals, and there
are chaotic dust lanes extending outward to 7.5 kpc. The fine-structure comparison of H$\alpha$ surface brightness and
\oiii/H$\alpha$ indicator of the ionization parameter shows mismatches in the extended northern and southern emission filaments.
There are three candidate luminous star clusters, all located far from the emission-line filaments and dust lanes.

{\bf NGC 5252} has long been perhaps the best-known example of an AGN with ionization cones. This edge-on S0
system has a half-spiral of dust (possibly only the front half of a more complete inner disk), whose orientation is consistent
with the kinematics of the ionized gas but not the stellar disk. As shown previously in both optical emission and H I
data, the gas kinematics are strongly decoupled from the stars. The gas, presumably externally acquired, has
internally well-ordered kinematics; in our Fabry-Perot maps, local regions of $\simeq 2${\arcsec} extent differ from
a smooth velocity field by no more than $\sigma_v = 27$ km s$^{-1}$. We find no mismatch between H$\alpha$ surface brightness
and ionization parameter in the outer filaments, suggesting that the WFPC2 images resolved most of the density contrast here;
the inner filaments are too small for this statistical test. The derived abundances show a substatial gradient from the near-nuclear
features to the outer filaments, which could indicate that both a central outflow and less-processed tidal mateiral are present.

{\bf NGC 5972} combines low redshift with intrinsically large EELRs detected to at least 50-kpc radius, revealing very rich structure. 
The starlight shows tidal tails and shells, crossed by dust lanes which span a $66^\circ$ range in orientation. These dust lanes
can be fit schematically by a differentially precessing, warped disk with a structured radial distribution; such a distribution is
seen in models with ages $\sim 1.5$ Gyr. The gas velocity field includes asymmetric or multiple line profiles north and south
the nucleus, consistent with seeing into a patchy distribution of dust so the data sample different depths on adjacent lines of sight.
The velocity field shows local departures $\sigma_v < 39$ km s$^{-1}$ from a smoothed distribution; these departures are 
broadly correlated with dust lanes, indicating a role for line-of-sight weighting. 
There is only a partial match between the dust lanes and ionized gas in front of the central bulge; projection effects and illumination of
the gas only within ionization cones seen nearly side-on could explain this. An \oiii/H$\alpha$ comparison shows evidence for
unresolved density structure in the inner emission regions but not the outer areas. There is a modest population of candidate star
clusters, not spatially associated with the dust lanes or ionized-gas filaments. The outermost filaments of ionized gas align roughly with
stellar tails, but their centerlines differ, suggesting either star/gas sorting as a result of different initial velocity distributions during the
interaction or three-dimensional differences highlighted by the AGN's radiation pattern. The velocity field, both sampled wih a long slit and via Fabry--Perot mapping,
shows ripples where dust attentuation is important along the major axis, broadly consistent with the warped-disk  geometry found from the dust lanes themselves.

{\bf SDSS 1510+07} is an SB0 galaxy with a faint tail connecting it to a dim companion at projected distance 45{\arcsec} (40 kpc), and extending
beyond that companion. Its EELRs consist of two large, filamentary clouds roughly along the galaxy's minor axis, with some coherent 
filaments stretching as much as 7.5 kpc. These structures follow the expected relation between ionization level and H$\alpha$
surface brightness, suggesting that the HST data resolve most of the density contrast in them.

{\bf SDSS 2201+11} has a continuum morphology consistent with a pure bulge plus absorption in a warped (differentially precessing)
disk. The EELRs are two distinct clouds, largely disjoint from the emission around the AGN,
 about 20$^\circ$ out of the midplane of the dust and starlight, with a dust lane connecting to the northern cloud
showing it to be on the near side (and receding, from the velocity data). The inner parts of the starlight distribution
share the rotation sense of the gas. Deep images show outer tidal loops. The southern cloud may
have two kinematically separate components. The dust lanes are well understood as parts of a precessing disk of material
which would have been externally acquired $\sim 1.7$ Gyr ago; if they are part of this disk, the EELR clouds may span a much wider
cone angle than we observe. The gas velocity field has scatter among observed regions $\sigma_v < 59$ km s$^{-1}$
about a smooth model. There may be a few luminous star clusters, but the candidates are too close to the dust lanes for accurate photometry.
Both clouds show evidence of unresolved fine-structure, through mismatches between the observed and predicted relations between
H$\alpha$ surface brightness and \oiii/H$\alpha$.  The abundances are substantially below solar, consistent with an origin of the cloud material
in tidal distortion of a low-mass, low-abundance former companion.

{\bf The Teacup AGN} has an EELR dominated by a 13-kpc filamentary loop to the NW. The continuum shows an early-type morphology with
a probable shell plus an oppositely directed tail, as seen in simulation of minor mergers involving a small, dynamically cold intruder.
There are a few candidate luminous star clusters; a clump of such candidates half-behind the gas loop may be a background disk galaxy
with redder core. Chaotic dust lanes appear up to 2{\arcsec} (3 kpc) from the center, adding to the evidence for a past merger.
H$\alpha$ and \oiii/H$\alpha$ pixel values suggest unresolved structure at increasing levels for projected distances $r>3.3${\arcsec} (5 kpc) from the AGN, becoming
more pronounced in the loop at  $r>5$\arcsec (8 kpc). Recent results by \cite{Harrson} show strongly blueshifted line emission just SE of the AGN,
and a radio-continuum counterpart to the large emission loop to thenortheast, suggesting outflow on these two scales. Near-solar abundances in
the gas would be broadly consistent with an important role for AGN-driven outflow.

{\bf UGC 7342} shows a two-sided, roughly conical set of emission filaments extending at least 37 kpc from the AGN. The host galaxy shows complex
dust lanes and tidal tails. The apparent companion galaxy SDSS J121820.18+291447.7 lies in the background at $z=0.0674$,, so the merger is advanced 
enough for the second galaxy to no longer
be distinct. This system has a rich set of candidate luminous star clusters, with the brightest close to the center line of the brightest tidal tail.
The gas velocity field shows ordered rotation, with significant ripples crossing the tidal tails. Some ionized-gas filaments roughly match stellar tidal tails, but others have no continuum counterparts, so these distributions
are largely uncoupled. The inner filaments in this system show a mismatch with predictions of the H$\alpha$ -- \oiii/H$\alpha$ relation,
indicating substantial levels of unresolved density structure. We find low abundances througout the EELR in this system, suggesting, with the 
morphological data, that it consists of tidal debris from a low-abundance disrupted system. 

{\bf UGC 11185} is part of a strongly interacting galaxy pair, with a prominent tidal loop to the east and chaotic dust structure. Two of the four candidate
star clusters are found together in this loop. The emission-line data suggest that the HST images resolve most of the density structure in the gas throughout the
EELR, which plausibly traces a large set of ionization cones. The longslit data show multiple velocity compones in \oiii to the east of the nucleus, with the strongest  associated morphogicaly with the triangular or conical bright regions seen in the HST \oiii image. Along with the relatively high abundances, this suggests that
a nuclear outflow has been important in shaping the EELR of this system.

\clearpage

\end{document}